\documentclass[article, nojss]{jss}

\usepackage{amsfonts,amsmath}

\author{Dimitris Rizopoulos\\ Erasmus Medical Center Rotterdam}

\title{The \proglang{R} Package \pkg{JMbayes} for Fitting Joint Models for Longitudinal 
and Time-to-Event Data using MCMC}

\Plainauthor{Dimitris Rizopoulos}
\Plaintitle{The R Package JMbayes for Fitting Joint Models for Longitudinal 
and Time-to-Event Data using MCMC}
\Shorttitle{R package JMbayes}

\Abstract{
  Joint models for longitudinal and time-to-event data constitute an attractive modeling 
  framework that has received a lot of interest in the recent years. This paper presents 
  the capabilities of the \proglang{R} package \pkg{JMbayes} for fitting these models 
  under a Bayesian approach using Markon chain Monte Carlo algorithms. \pkg{JMbayes} can 
  fit a wide range of joint models, including among others joint models for continuous and
  categorical longitudinal responses, and provides several options for modeling the 
  association structure between the two outcomes. In addition, this package can be used to
  derive dynamic predictions for both outcomes, and offers several tools to validate these
  predictions in terms of discrimination and calibration. All these features are 
  illustrated using a real data example on patients with primary biliary cirrhosis.
}
\Plainkeywords{survival analysis, time-varying covariates, random effects, mixed models, 
dynamic predictions, validation}

\Address{
  Dimitris Rizopoulos\\
  Department of Biostatistics\\
  Erasmus Medical Center Rotterdam\\
  PO Box 2040, 3000 CA Rotterdam, the Netherlands\\
  E-mail: \email{d.rizopoulos@erasmusmc.nl}\\
  URL: \url{http://www.erasmusmc.nl/biostatistiek/People/Faculty/drizopoulos/}
}

\newcommand{\bfalpha}{\mbox{{\boldmath $\alpha$}}}
\newcommand{\bfbeta}{\mbox{{\boldmath $\beta$}}}
\newcommand{\bfgamma}{\mbox{{\boldmath $\gamma$}}}
\newcommand{\bftheta}{\mbox{{\boldmath $\theta$}}}

\newcommand{\bw}{\mbox{{\boldmath $w$}}}
\newcommand{\bv}{\mbox{{\boldmath $v$}}}
\newcommand{\by}{\mbox{{\boldmath $y$}}}
\newcommand{\bx}{\mbox{{\boldmath $x$}}}
\newcommand{\bX}{\mbox{{\boldmath $X$}}}

\newcommand{\bz}{\mbox{{\boldmath $z$}}}
\newcommand{\bZ}{\mbox{{\boldmath $Z$}}}
\newcommand{\bK}{\mbox{{\boldmath $K$}}}
\newcommand{\bb}{\mbox{{\boldmath $b$}}}

\newcommand{\bD}{\mbox{{\boldmath $D$}}}
\newcommand{\aucHat}{A$\widehat{\mbox{U}}$C}
\newcommand{\ipeHat}{I$\widehat{\mbox{P}}$E}

\begin{document}



\section[Introduction]{Introduction} \label{Sec:Intro}
Joint models for longitudinal and time-to-event data constitute an attractive modeling 
paradigm that currently enjoys great interest in the statistics and medical literature 
\citep{rizopoulos.lesaffre:14, rizopoulos:12, tsiatis.davidian:04}. These models are 
utilized in follow-up studies where interest is in associating a longitudinal response
with an event time outcome. In general, there are mainly two settings in which such type 
of models are required. First, when one is interested in measuring the strength of the 
association between the hazard of an event and a time-varying covariate, then we should
pay special attention to the attributes of the covariate process. In particular, when this
is an endogenous time-varying covariate \citep[Section~6.3]{kalbfleisch.prentice:02}, 
standard methods, such as the time-dependent Cox model \citep{therneau.grambsch:00}, are 
not optimal for measuring this association. Standard examples of endogenous covariates are
covariates, which are measured on the sample units themselves, for instance, biomarkers or
other parameters measured on patients during follow-up. The important feature of such 
covariates is that their existence and/or future path is directly related to the event 
status. By postulating a model for the joint distribution of the covariate and event 
processes we explicitly acknowledge this link, and hence we obtain a more accurate 
estimate for their association. The second case in which joint models are of use is when 
one needs to account for incomplete data. More specifically, when the probability of 
missingness depends on unobserved longitudinal responses, then in order to obtain valid 
inferences we need to postulate a model for the joint distribution of the longitudinal and
missingness processes \citep{little.rubin:02, molenberghs.kenward:07}. In this context, 
three main frameworks have been proposed to define such joint distributions, namely, 
selection, pattern mixture and shared parameter models. The majority of the models
that have been proposed in the literature under these frameworks have focused on standard 
designs assuming a fixed set of time points at which subjects are expected to provide 
measurements. Nonetheless, in reality, subjects often do not adhere to the posited study 
schedule and may skip visits and dropout from the study at random time points. Even though
in many of those occasions information on the exact dropout time is available, the typical
convention in selection and pattern mixture modeling has been to ignore this feature and 
coerce measurements back to discrete follow-up times. Another alternative that makes 
better use of the available data is to acknowledge that dropout occurs in continuous time 
and consider it as a time-to-event outcome.

Following the increasing in these models, currently there are several software 
implementations available to fit them. The \proglang{R} package \pkg{JM} \citet{JM, 
rizopoulos:12, rizopoulos:10} fits joint models for a continuous longitudinal outcome and 
an event time process under maximum likelihood. Several types of association structures 
are supported and the package also allows to fit joint models with competing risk survival
data. In addition, \pkg{JM} can be used to calculate dynamic predictions for either of the
two outcomes. The \proglang{R} package \pkg{joineR} \citep{joineR} similarly fits joint 
models for a continuous longitudinal outcome and a time-to-event, following the 
formulation of \citet{henderson.et.al:00}. In addition, the \pkg{stjm} package for 
\proglang{STATA} \citep{STJM} implements joint modeling of a normal longitudinal response 
and a time-to-event using maximum likelihood, with emphasis on parametric time-to-event 
models. The implementation of joint models in \proglang{SAS} and \pkg{WinBUGS} has been 
discussed by \citet{guo.carlin:04}. Finally, contrary to the previous software 
implementations, function \code{Jointlcmm()} from the \proglang{R} package \pkg{lcmm} 
\citep{lcmm} fits joint latent class models for a continuous longitudinal outcome and a 
survival outcome using maximum likelihood; these models postulate that the association 
between the two processes is captured by categorical random effects (i.e., latent 
classes). 

In this paper we introduce the \proglang{R} package \pkg{JMbayes} that fits joint models
under a Bayesian approach. \pkg{JMbayes} can fit a wide range of joint models, including
among others joint models for continuous and categorical longitudinal responses. It 
provides several options for modeling the association structure between the two outcomes,
with the possibility of different terms from the longitudinal submodel entering the linear
predictor of the survival submodel, allowing also for general transformation of these 
terms. In addition, the package provides extensive capabilities to derive dynamic 
predictions for both outcomes, it allows to combine predictions from different models 
using innovative Bayesian model averaging techniques, and facilitates the utilization of 
these predictions in practice using a web interface. Moreover, it offers several tools to 
quantify the quality of these predictions in terms of discrimination and calibration, and 
code is provided for their validation. The rest of the paper is organised as follows. 
Section~\ref{Sec:Theory} presents a short review of the underlying methodological 
framework behind joint models. Section~\ref{Sec:pkgJMbayes} gives the details behind the 
implementation of joint models in package \pkg{JMbayes}, and 
Section~\ref{Sec:JMbayesBasicUse} illustrates in detail the use of the package in a real 
dataset on patients with primary biliary cirrhosis. Finally, Section~\ref{Sec:DynPred} 
presents how dynamic predictions for the longitudinal and event time outcomes are defined, 
and how they can be calculated and validated with the package.


\section[Theoretical framework]{Theoretical framework} \label{Sec:Theory}
Let $\mathcal D_n = \{T_i, \delta_i, \by_i; i = 1, \ldots, n\}$ denote a sample from the 
target population, where $T_i^*$ denotes the true event time for the $i$-th subject , 
$C_i$ the censoring time, $T_i = \min(T_i^*, C_i)$ the corresponding observed event time, 
and $\delta_i = I(T_i^* \leq C_i)$ the event indicator, with $I(\cdot)$ being the 
indicator function that takes the value 1 when $T_i^* \leq C_i$, and 0 otherwise. 
In addition, we let $\by_i$ denote the $n_i \times 1$ longitudinal response vector for the
$i$-th subject, with element $y_{il}$ denoting the value of the longitudinal outcome taken
at time point $t_{il}$, $l = 1, \ldots, n_i$.

To accommodate different types of longitudinal responses in a unified framework, we 
postulate a generalized linear mixed effects model. In particular, the conditional 
distribution of $\by_i$ given a vector of random effects $\bb_i$ is assumed to be a member
of the exponential family, with linear predictor given by
\begin{equation}
g \bigl [ E \{ y_i(t) \mid \bb_i \} \bigr ] = \eta_i(t) = \bx_i^\top(t) \bfbeta + 
\bz_i^\top(t) \bb_i, \label{Eq:MixedModel}
\end{equation}
where $g(\cdot)$ denotes a known one-to-one monotonic link function, and $y_i(t)$ 
denotes the value of the longitudinal outcome for the $i$-th subject at time point $t$, 
$\bx_i(t)$ and $\bz_i(t)$ denote the time-dependent design vectors for the
fixed-effects $\bfbeta$ and for the random effects $\bb_i$, respectively. The random 
effects are assumed to follow a multivariate normal distribution with mean zero and 
variance-covariance matrix $\bD$. For the survival process, we assume that the risk for an
event depends on a function of the subject-specific linear predictor $\eta_i(t)$. More 
specifically, we have
\begin{eqnarray}
\nonumber h_i (t \mid \mathcal H_i(t), \bw_i) & = & \lim_{\Delta t \rightarrow 0} 
\frac{1}{\Delta t}\Pr \{ t \leq T_i^* < t + \Delta t \mid T_i^* \geq t, \mathcal H_i(t), 
\bw_i \} \\
& = & h_0(t) \exp \bigl  [\bfgamma^\top
\bw_i + f \{\eta_i(t), \bb_i, \bfalpha \} \bigr] , \quad t > 0, \label{Eq:Surv-RR}
\end{eqnarray}
where $\mathcal H_i(t) = \{ \eta_i(s), 0 \leq s < t \}$ denotes the history of the 
underlying longitudinal process up to $t$, $h_0(\cdot)$ denotes the baseline hazard 
function, $\bw_i$ is a vector of baseline covariates with corresponding 
regression coefficients $\bfgamma$. Parameter vector $\bfalpha$ quantifies the association
between features of the marker process up to time $t$ and the hazard for an event at the 
same time point. Various options for the form of function $f(\cdot)$ are presented in 
Section~\ref{Sec:JMbayesBasicUse-Ass}. To complete the specification of the survival 
process we need to make appropriate assumptions for the baseline hazard function 
$h_0(\cdot)$. To model this function, while still allowing for flexibility, we use a 
B-splines approach. In particular, the logarithm of the  baseline hazard function is 
expressed as
\begin{equation}
\log h_0(t) = \gamma_{h_0,0} + \sum \limits_{q = 1}^Q \gamma_{h_0,q} B_q(t, \bv),
\label{Eq:BaseHaz}
\end{equation}
where $B_q(t, \bv)$ denotes the $q$-th basis function of a B-spline with knots 
$v_1, \ldots, v_Q$ and $\bfgamma_{h_0}$ the vector of spline coefficients. Increasing the 
number of knots $Q$ increases the flexibility in approximating $\log h_0(\cdot)$; however,
we should balance bias and variance and avoid over-fitting. A standard rule of thumb is to 
keep the total number of parameters, including the parameters in the linear predictor in 
\eqref{Eq:Surv-RR} and in the model for $h_0(\cdot)$, between 1/10 and 1/20 of the total 
number of events in the sample \citep[Section~4.4]{harrell:01}. After the number of knots 
has been decided, their location can be based on percentiles of the observed event times 
$T_i$ or of the true event times $\{T_i : T_i^* \leq C_i, i = 1, \ldots, n\}$ in order to 
allow for more flexibility in the region of greatest density. A standard alternative 
approach that avoids the task of choosing the appropriate number and position of the knots
is to include a relatively high number of knots (e.g., 15 to 20) and appropriately 
penalize the B-spline regression coefficients $\bfgamma_{h_0}$ for smoothness 
\citep{eilers.marx:96}.

Under the Bayesian approach, estimation of joint model's parameters proceeds using Markov 
chain Monte Carlo (MCMC) algorithms. The expression for the posterior distribution of the 
model parameters is derived under the assumptions that given the random effects, both the 
longitudinal and event time process are assumed independent, and the longitudinal 
responses of each subject are assumed independent. Formally we have,
\begin{eqnarray}
p(\by_i, T_i, \delta_i \mid \bb_i, \bftheta) & = & p(\by_i \mid \bb_i, \bftheta) \; 
p(T_i, \delta_i \mid \bb_i, \bftheta), \label{Eq:CondInd-I}\\
p(\by_i \mid \bb_i, \bftheta) & = & \prod_l p ( y_{il} \mid \bb_i, \bftheta ), 
\label{Eq:CondInd-II}
\end{eqnarray}
where $\bftheta$ denotes the full parameter vector, and $p(\cdot)$ denotes an appropriate 
probability density function. Under these assumptions the posterior distribution is 
analogous to:
\begin{eqnarray}
p(\bftheta, \bb) \propto \prod \limits_{i = 1}^n \prod \limits_{l = 1}^{n_i} 
p (y_{il} \mid \bb_i, \bftheta) \; p(T_i, \delta_i \mid \bb_i, \bftheta) \; 
p(\bb_i, \bftheta) \; p(\bftheta), \label{Eq:FullPost}
\end{eqnarray}
where
\[
p (y_{il} \mid \bb_i, \bftheta) = \exp \biggl \{ \Bigl [y_{il} \psi_{il}(\bb_i) - 
c\{\psi_{il}(\bb_i)\} \Bigr ] \Big / a(\varphi) - d(y_{il}, \varphi) \biggr \},
\]
with $\psi_{il}(\bb_i)$ and $\varphi$ denoting the natural and dispersion parameters in the 
exponential family, respectively, $c(\cdot)$, $a(\cdot)$, and $d(\cdot)$ are known 
functions specifying the member of the exponential family, and for the survival part
\[
p(T_i, \delta_i \mid \bb_i, \bftheta) = h_i(T_i \mid \mathcal H_i(T_i))^{\delta_i} 
\exp \Bigl \{- \int_0^{T_i} h_i(s \mid \mathcal H_i(s) ) \; ds \Bigl\},
\]
with $h_i(\cdot)$ given by \eqref{Eq:Surv-RR}. The integral in the definition of the 
survival function
\begin{equation}
S_i(t \mid \mathcal H_i(t), \bw_i) = \exp \Bigl \{- \int_0^t h_0(s) 
\exp \bigl  [\bfgamma^\top \bw_i + f \{\eta_i(s), \bfalpha \} \bigr] ds \Bigr \},
\label{Eq:SurvivalFun}
\end{equation}
does not have a closed-form solution, and thus a numerical method must be employed for its
evaluation. Standard options are the Gauss-Kronrod and Gauss-Legendre quadrature rules. 

For the parameters $\bftheta$ we take standard prior distributions. In particular, for the
vector of fixed effects of the longitudinal submodel $\bfbeta$, for the regression 
parameters of the survival model $\bfgamma$, for the vector of spline coefficients for the
baseline hazard $\bfgamma_{h_0}$, and for the association parameter $\bfalpha$ we use 
independent univariate diffuse normal priors. The penalized version of the B-spline 
approximation to the baseline hazard can be fitted by specifying for $\bfgamma_{h_0}$ the 
improper prior \citep{lang.brezger:04}:
\[
p(\bfgamma_{h_0} \mid \tau_h) \propto \tau_h^{\rho(K)/2}\exp \Bigl (-\frac{\tau_{h}}{2} 
\bfgamma_{h_0}^\top \bK \bfgamma_{h_0} \Bigr ),
\]
where $\tau_h$ is the smoothing parameter that takes a $\mbox{Gamma}(1, 0.005)$ 
hyper-prior in order to ensure a proper posterior for $\bfgamma_{h_0}$, $\bK = 
\Delta_r^\top \Delta_r$, where $\Delta_r$ denotes $r$-th difference penalty matrix, and 
$\rho(\bK)$ denotes the rank of $\bK$. For the covariance matrix of the random effects we 
assume an inverse Wishart prior, and when fitting a joint model with a normally 
distributed longitudinal outcome, we take an inverse-Gamma prior for the variance of the 
error terms $\sigma^2$. More details regarding Bayesian estimation of joint models can be 
found in \citet[Chapter~7]{ibrahim.et.al:01} and \citet{Brown.et.al:05}.


\section[The R package JMbayes]{The \proglang{R} package \pkg{JMbayes}} 
\label{Sec:pkgJMbayes}
\subsection[Design]{Design} \label{Sec:pkgJMbayes-Design}
In many regards the design of package \pkg{JMbayes} is similar to the one of package 
\pkg{JM} for fitting joint models under maximum likelihood. In particular, \pkg{JMbayes} 
has a basic model-fitting function called \code{jointModelBayes()}, which accepts as main 
arguments a linear mixed effects object fit as returned by functions \code{lme()} of 
package \pkg{nlme} \citep{nlme} or from function \code{glmmPQL()} from package \pkg{MASS} 
\citep{MASS}, and a survival object fit as returned by function \code{coxph()} of package 
\pkg{survival} \citep{survival}. The final required argument is \code{timeVar}, a 
character string that denotes the name of the time variable in the mixed model. By default
\code{jointModelBayes()} fits joint models with a linear mixed effects submodel for a 
continuous longitudinal outcome, and a relative risk submodel of the form 
\eqref{Eq:Surv-RR} with $f \{\eta_i(t), \bb_i, \bfalpha \} = \alpha \eta_i(t)$, i.e., the 
risk for an event at time $t$ is associated with the subject-specific mean of the 
longitudinal outcome at the same time point. Joint models for other types of longitudinal 
outcomes can be fitted by appropriately specifying argument \code{densLong}, and arguments
\code{param}, \code{extraForm} and \code{transFun} can be used to add extra terms 
involving components of the longitudinal process and possibly transform these terms. A 
detailed account on the use of these arguments, with examples, is given in 
Section~\ref{Sec:JMbayesBasicUse}. The baseline hazard is by default approximated using 
penalized B-splines; regression splines can be instead invoked by appropriately setting 
argument \code{baseHaz}. The number and position of the knots can be controlled via the 
\code{lng.in.kn} and \code{knots} control arguments. The former defines the number of 
knots to use (by default placed at equally spaced percentiles of the observed event times),
whereas argument \code{knots} can be invoked to specify knots at specific positions. The 
type of numerical integration algorithm used to approximate the survival function 
\eqref{Eq:SurvivalFun} is specified with the control argument \code{GQsurv} with options
\code{"GaussKronrod"} (default) and \code{"GaussLegendre"}, while the number of quadrature
points is specified using the control argument \code{GQsurv.k} (for the Gauss-Kronrod rule
only 7 or 15 can be specified). The fitting process can be further appropriately tweaked 
using a series of extra control arguments explained in the following section and in the
help file of \code{jointModelBayes()}. In addition, the default values of the parameters 
of the prior distributions can be altered using the \code{priors} argument, and 
analogously the default initial values using argument \code{init}.


\subsection[Implementation details]{Implementation details} \label{Sec:pkgJMbayes-Impl}
The MCMC algorithm that samples from the posterior conditional distributions of the 
parameters and the random effects is implemented by the internal function 
\code{MCMCfit()}. For the majority of the posterior conditionals random walk Metropolis is
used, with exceptions for the precision parameter of the error terms distribution when a 
linear mixed model is used for the longitudinal outcome in which case slice sampling is 
used, and for the random effects precision matrix $\bD^{-1}$ in which case when the random
effects are assumed normally distributed the posterior conditional is a Wishart 
distribution (if argument \code{df.RE} of \code{jointModelBayes()} is not 
\code{NULL} the distribution of the random effects is assumed to be a Student's-$t$ 
distribution with \code{df.RE} degrees of freedom; in this case the random effects 
precision matrix is updated with a Metropolis-Hastings algorithm). The implementation 
behind \code{MCMCfit()} takes full advantage of the separately fitted mixed effects and 
Cox models in order to appropriately define the covariance matrix of the normal proposal 
distributions for the random walk Metropolis algorithm. In particular, for $\bfbeta$ and 
$\bb_i$ these covariance matrices are taken from the mixed model, whereas for the 
regression coefficients in the linear predictor of the survival submodel and the B-spline 
coefficients $\bfgamma_{h_0}$ a two-stage approach is employed, where a 
time-dependent Cox model is fitted using the mixed model to compute $f \{\eta_i(t), 
\bb_i, \bfalpha \}$. These proposal distributions are tuned during an adaptive phase of
\code{n.adapt} iterations (default 3000), where every \code{n.batch} iterations (default 
100) the acceptance rate of the algorithms are checked. Following a burn-in period of 
\code{n.burnin} iterations (default 3000) is performed, and after these iterations the 
algorithm continues to run for an extra of \code{n.iter} iterations (default 20000). 
The chains are thinned according to the \code{n.thin} argument (default is to keep 2000 
iterations for each parameter).

From the two schools of running MCMC algorithms, namely the `one long chain school' and
the `multiple shorter chains school', \pkg{JMbayes} implements the former. Users who wish 
to check convergence using multiple chains can still do it but with a bit of extra 
programming. More specifically, they could call \code{jointModelBayes()} with different 
initial values (by appropriately specifying argument \code{init}), and following they 
could extract component \code{mcmc} from the fitted models, which is the list of simulated 
values for each parameter. These lists could subsequently be processed using the 
\pkg{coda} package \citep{coda} and perform these diagnostic tests.


\section[Practical use of JMbayes]{Practical use of \pkg{JMbayes}} \label{Sec:JMbayesBasicUse}
\subsection[The basic joint model]{The basic joint model} \label{Sec:JMbayesBasicUse-Std}
We will illustrate the capabilities of package \pkg{JMbayes} using the primary biliary
cirrhosis (PBC) data collected by the Mayo Clinic from 1974 to 1984 
\citep{Murtaugh.et.al:94}. PBC is a chronic, fatal, but rare liver disease characterized 
by inflammatory destruction of the small bile ducts within the liver, which eventually 
leads to cirrhosis of the liver. Patients with PBC have abnormalities in several blood 
tests, such as elevated levels of serum bilirubin. For our analysis we will consider 312 
patients who have been randomized to D-penicillamine and 154 placebo. During follow-up 
several biomarkers associated with PBC have been collected for these patients. Here we 
focus on serum bilirubin levels, which is considered one of the most important ones 
associated with disease progression. Patients had on average 6.2 
measurements (std. deviation 3.8 measurements), with a total of 1945 
observations. In package \pkg{JMbayes} the PBC data are available in the data frames 
\code{pbc2} and \code{pbc2.id} containing the longitudinal and survival information, 
respectively (i.e., the former is in the long format while the latter contains a single 
row per patient).

\begin{figure}
\centering
\includegraphics[width = 0.9\textwidth]{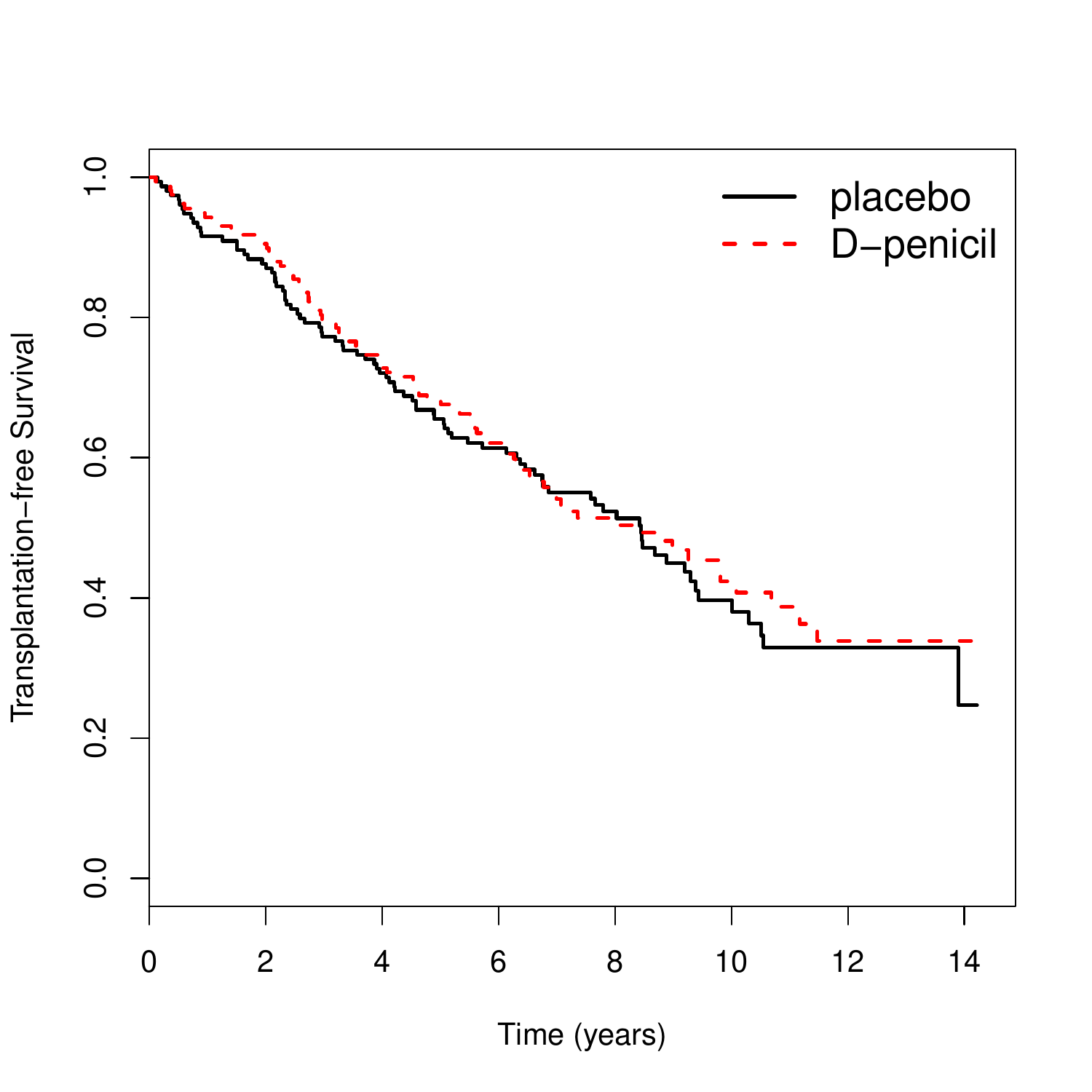}
\caption{Kaplan-Meier estimator of transplantation-free survival 
probabilities for the two treatment groups.} \label{Fig:KM}
\end{figure}

\begin{figure}
\centering
\includegraphics[width = 0.9\textwidth]{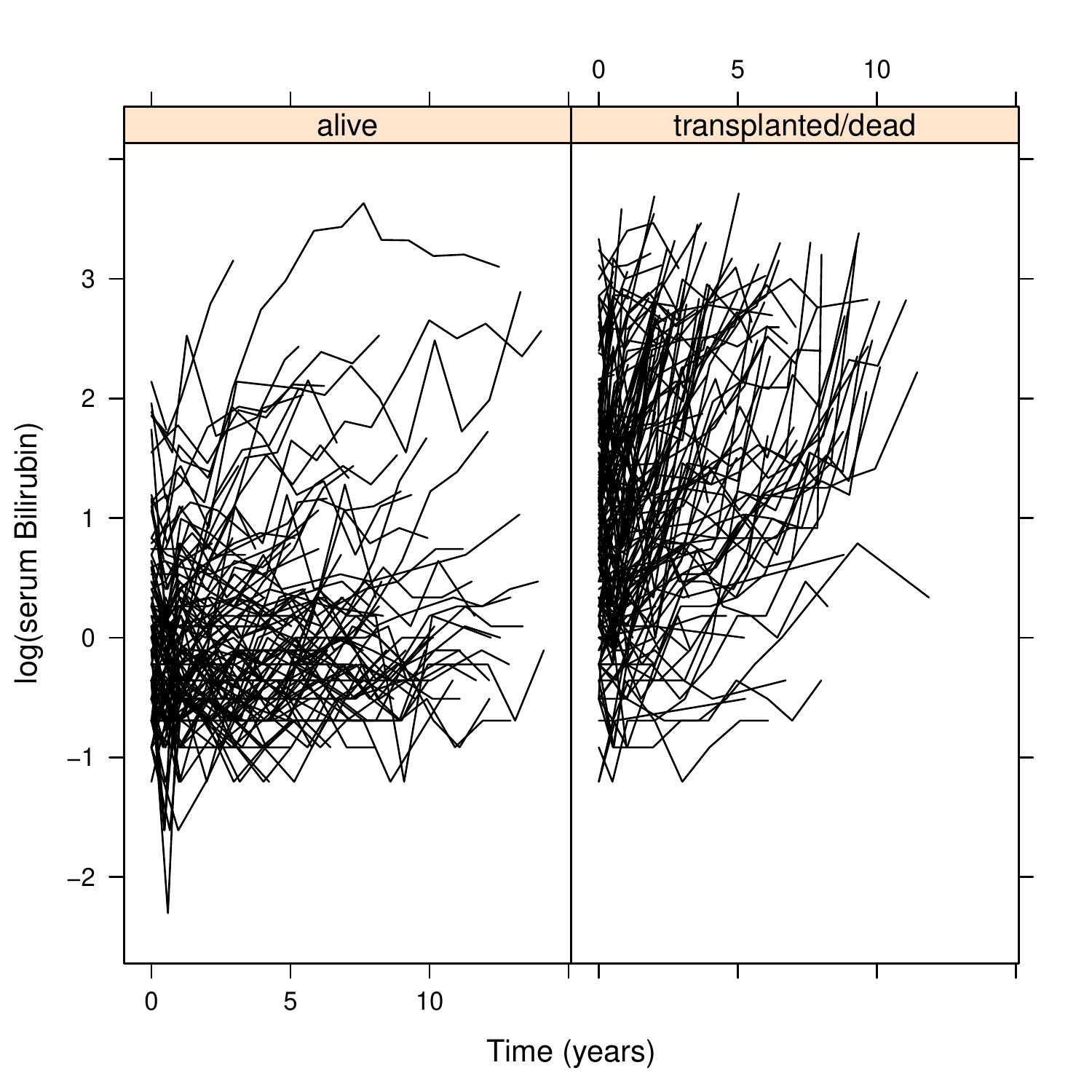}
\caption{Subject-specific longitudinal trajectories for log 
serum bilirubin for patients with and without an event.} \label{Fig:SubjProfs}
\end{figure}

We start by loading packages \pkg{JMbayes} and \pkg{lattice} \citep{lattice} and defining 
the indicator \code{status2} for the composite event, namely transplantation or death: 
\begin{Schunk}
\begin{Sinput}
R> library("JMbayes")
R> library("lattice")
R> pbc2$status2 <- as.numeric(pbc2$status != "alive")
R> pbc2.id$status2 <- as.numeric(pbc2.id$status != "alive")
\end{Sinput}
\end{Schunk}
Descriptive plots for the survival and longitudinal outcomes are presented in 
Figures~\ref{Fig:KM} and \ref{Fig:SubjProfs} that depict the Kaplan-Meier estimate of 
transplantation-free survival for the two treatment groups, and the sample 
subject-specific longitudinal trajectories for patients with and without and endpoint, 
respectively.
\begin{Schunk}
\begin{Sinput}
R> sfit <- survfit(Surv(years, status2) ~ drug, data = pbc2.id)
R> plot(sfit, lty = 1:2, lwd = 2, col = 1:2, mark.time = FALSE, 
+       xlab = "Time (years)", ylab = "Transplantation-free Survival")
R> legend("topright", levels(pbc2.id$drug), lty = 1:2, col = 1:2, lwd = 2, 
+         cex = 1.3, bty = "n")
\end{Sinput}
\end{Schunk}
\begin{Schunk}
\begin{Sinput}
R> pbc2$status2f <- factor(pbc2$status2, levels = 0:1, 
+                          labels = c("alive", "transplanted/dead"))
R> xyplot(log(serBilir) ~ year | status2f, group = id, data = pbc2, type = "l", 
+         col = 1, xlab = "Time (years)", ylab = "log(serum Bilirubin)")
\end{Sinput}
\end{Schunk}

We continue by separately fitting a linear mixed model for the longitudinal and a Cox 
model for the survival one. Careful investigation of the shapes of the log 
serum bilirubin profiles indicates that for many individuals these seem to be nonlinear. 
Hence, to allow for flexibility in the specification of these profiles we include natural
cubic splines in both the fixed- and random-effects parts of the mixed model. This model
can be fitted using the following call to functions \code{lme()} and \code{ns()} (the 
latter from package \pkg{splines}):
\begin{Schunk}
\begin{Sinput}
R> lmeFit.pbc1 <- lme(log(serBilir) ~ ns(year, 2), data = pbc2,
+                     random = ~ ns(year, 2) | id)
\end{Sinput}
\end{Schunk}
Analogously, in the Cox model we control for treatment and age, and also allow for their 
interaction:
\begin{Schunk}
\begin{Sinput}
R> coxFit.pbc1 <- coxph(Surv(years, status2) ~ drug * age, data = pbc2.id, 
+                       x = TRUE)
\end{Sinput}
\end{Schunk}
In the call to \code{coxph()} argument \code{x} is set to \code{TRUE} such that the design
matrix is also included in the resulting model object. Using as main arguments the 
\code{lmeFit.pbc1} and \code{coxFit.pbc1} objects, the corresponding joint model is fitted
using the code:
\begin{Schunk}
\begin{Sinput}
R> jointFit.pbc1 <- jointModelBayes(lmeFit.pbc1, coxFit.pbc1, timeVar = "year", 
+                                   n.iter = 30000)
R> summary(jointFit.pbc1)
\end{Sinput}
\end{Schunk}
\begin{Schunk}
\begin{Soutput}
Call:
jointModelBayes(lmeObject = lmeFit.pbc1, survObject = coxFit.pbc1, 
    timeVar = "year", n.iter = 30000)

Data Descriptives:
Longitudinal Process		Event Process
Number of Observations: 1945	Number of Events: 169 (54.2%)
Number of subjects: 312

Joint Model Summary:
Longitudinal Process: Linear mixed-effects model
Event Process: Relative risk model with penalized-spline-approximated 
		baseline risk function
Parameterization: Time-dependent value 

      LPML      DIC      pD
 -3168.647 6103.934 931.971

Variance Components:
              StdDev    Corr        
(Intercept)   1.0067  (Intr)  n(,2)1
ns(year, 2)1  2.3131  0.3482        
ns(year, 2)2  2.2224  0.3250  0.5457
Residual      0.3020                

Coefficients:
Longitudinal Process
              Value Std.Err Std.Dev   2.5%  97.5%      P
(Intercept)  0.4804  0.0100  0.0578 0.3674 0.5842 <0.001
ns(year, 2)1 2.3183  0.0244  0.1374 2.0174 2.5690 <0.001
ns(year, 2)2 2.2397  0.0331  0.1969 1.8420 2.6248 <0.001

Event Process
                     Value Std.Err  Std.Dev    2.5%    97.5%      P
drugD-penicil      -0.6746  0.0604   0.7769 -2.1489   0.8536  0.390
age                 0.0401  0.0017   0.0111  0.0197   0.0644 <0.001
drugD-penicil:age   0.0119  0.0011   0.0150 -0.0177   0.0400  0.432
Assoct              1.4132  0.0063   0.0974  1.2393   1.6193 <0.001
Bs.gammas1         -6.8276  0.0917   0.6772 -8.2563  -5.6207 <0.001
Bs.gammas2         -6.7614  0.0948   0.6473 -8.1468  -5.6014 <0.001
Bs.gammas3         -6.6807  0.0977   0.6381 -8.0805  -5.5619 <0.001
Bs.gammas4         -6.5969  0.0984   0.6340 -7.9645  -5.5084 <0.001
Bs.gammas5         -6.5218  0.0978   0.6343 -7.8877  -5.4386 <0.001
Bs.gammas6         -6.4483  0.0995   0.6392 -7.8177  -5.3393 <0.001
Bs.gammas7         -6.3792  0.1002   0.6423 -7.7613  -5.2463 <0.001
Bs.gammas8         -6.3245  0.0984   0.6358 -7.6902  -5.2296 <0.001
Bs.gammas9         -6.2744  0.0974   0.6344 -7.6178  -5.1546 <0.001
Bs.gammas10        -6.2387  0.0942   0.6274 -7.5891  -5.1369 <0.001
Bs.gammas11        -6.2109  0.0904   0.6204 -7.5382  -5.0980 <0.001
Bs.gammas12        -6.1989  0.0873   0.6183 -7.5206  -5.0971 <0.001
Bs.gammas13        -6.2086  0.0882   0.6239 -7.5497  -5.1103 <0.001
Bs.gammas14        -6.2396  0.0861   0.6418 -7.6330  -5.1410 <0.001
Bs.gammas15        -6.2783  0.0806   0.6765 -7.7393  -5.0614 <0.001
Bs.gammas16        -6.3173  0.0783   0.7381 -7.8507  -4.9740 <0.001
Bs.gammas17        -6.3405  0.0838   0.8348 -8.0335  -4.7474 <0.001
tauBs             268.0560 17.8397 201.6598 41.7713 786.8109     NA

MCMC summary:
iterations: 30000 
adapt: 3000 
burn-in: 3000 
thinning: 15 
time: 4 min
\end{Soutput}
\end{Schunk}
As explained earlier, argument \code{timeVar} is a character string that specifies the 
name of the time variable in the mixed model (the scale of time (e.g., days, months, 
years) in both the mixed and Cox models must be the same). In addition, using the control 
argument \code{n.iter} we specified that after adaption and burn-in, the MCMC should run 
for 30000 iterations. The default call to \code{jointModelBayes()} includes in the 
linear predictor of the relative risk model the 
subject-specific linear predictor of the mixed model $\eta_i(t)$, which in this case 
represents the average subject-specific log serum bilirubin level. The output of the 
\code{summary()} method is rather self-explanatory and contains model summary statistics,
namely LPML (the log pseudo marginal likelihood value), DIC (deviance information 
criterion), and pD (the effective number of parameters component of DIC), posterior means 
for all parameters, and standard errors (effective sample size estimated using time series
methodology), standard deviations, 95\% credibility intervals and tail probabilities for all 
regression coefficients in the two submodels. The association parameter $\alpha$ is 
denoted in the output as \code{Assoct}. The tail probabilities, under the column with the 
heading \code{P}, are calculated as $2\times \min\{\Pr(\theta > 0), \Pr(\theta < 0)\}$,
with $\theta$ denoting here the corresponding regression coefficient from the longitudinal
or the survival submodel. The results suggest that serum bilirubin is strongly associated 
with the risk for the composite event, with a doubling of serum bilirubin levels, 
resulting in a 2.7-fold (95\% CI: 2.4; 3.1) 
increase of the risk. In the appendix we show how the \code{plot()} method can be used 
produce diagnostic plots for investigating the convergence of the MCMC.

\newpage
\subsection[Extended joint models]{Extended joint models} \label{Sec:JMbayesBasicUse-Ext}
The previous section showed how the basic joint model for a continuous normally 
distributed longitudinal outcome and a time-to-event can be fitted in \pkg{JMbayes}. 
In this section we will illustrate how joint models for other types of longitudinal 
responses may be fitted using function \code{jointModelBayes()} by suitably specifying 
argument \code{densLong}. In particular, this argument accepts a function that calculates 
the probability density function (and its natural logarithm) of the longitudinal outcome, 
with arguments \code{y} denoting the vector of longitudinal responses $y$, \code{eta.y} 
the subject-specific linear predictor $\eta_i(t)$, \code{scale} a potential scale parameter 
(e.g., the standard deviation of the error terms), \code{log} a logical denoting whether
logarithm of the density is computed, and \code{data} a data frame that contains variables
that are potentially required in the definition of \code{densLong}. To better illustrate 
the use of this function, we present here three examples of joint models with more 
elaborate longitudinal outcomes. We start with an extension of model \code{jointFit.pbc1}
that allows for a more heavier-tailed error distribution, that is,
\begin{Schunk}
\begin{Sinput}
R> dLongST <- function (y, eta.y, scale, log = FALSE, data) {
+      dgt(x = y, mu = eta.y, sigma = scale, df = 4, log = log)
+  }
\end{Sinput}
\end{Schunk}
Function \code{dgt()} of package \pkg{JMbayes} calculates the probability density 
function of the generalized Student's t distribution (i.e., a Student's t with mean 
parameter \code{mu} and scale parameter \code{sigma}). Supplying this function in the 
\code{densLong} fits the corresponding joint model:
\begin{Schunk}
\begin{Sinput}
R> jointFit.pbc2 <- jointModelBayes(lmeFit.pbc1, coxFit.pbc1, timeVar = "year", 
+                                   densLong = dLongST)
R> summary(jointFit.pbc2)
\end{Sinput}
\end{Schunk}
\begin{Schunk}
\begin{Soutput}
. . .

Variance Components:
              StdDev    Corr        
(Intercept)   1.0092  (Intr)  n(,2)1
ns(year, 2)1  2.2591  0.3478        
ns(year, 2)2  1.9402  0.2780  0.5703
Residual      0.2087                

Coefficients:
Longitudinal Process
              Value Std.Err Std.Dev   2.5%  97.5%      P
(Intercept)  0.5175  0.0299  0.0780 0.3614 0.6368 <0.001
ns(year, 2)1 2.3095  0.0704  0.1865 1.9769 2.6235 <0.001
ns(year, 2)2 2.0617  0.0527  0.1832 1.7155 2.4198 <0.001

Event Process
                     Value Std.Err  Std.Dev    2.5%    97.5%      P
drugD-penicil      -0.9085  0.0688   0.7257 -2.4245   0.4983  0.200
age                 0.0357  0.0017   0.0096  0.0159   0.0537 <0.001
drugD-penicil:age   0.0162  0.0012   0.0138 -0.0100   0.0455  0.227
Assoct              1.3595  0.0060   0.0947  1.1827   1.5581 <0.001
Bs.gammas1         -6.5082  0.1027   0.5917 -7.6831  -5.3422 <0.001

. . .
\end{Soutput}
\end{Schunk}
We observe some slight changes in the regression coefficients of both submodels, where a 
doubling of serum bilirubin levels is now associated with a 2.6-fold 
(95\% CI: 2.3; 2.9) increase of the risk for the composite 
event.

Following we illustrate the use of \code{densLong} for fitting a joint model with a 
dichotomous (binary) longitudinal outcome. Since in the PBC data there was no binary
biomarker recorded during follow-up, we artificially create one by dichotomizing 
serum bilirubin at the threshold value of 1.8 mg/dL. To fit the corresponding joint model
we need first to fit a mixed effects logistic regression for the longitudinal binary 
outcome using function \code{glmmPQL()} from package \pkg{MASS}, the syntax is
\begin{Schunk}
\begin{Sinput}
R> pbc2$serBilirD <- as.numeric(pbc2$serBilir > 1.8)
R> lmeFit.pbc2 <- glmmPQL(serBilirD ~ year, random = ~ year | id,
+                         family = binomial, data = pbc2)
\end{Sinput}
\end{Schunk}
As for continuous longitudinal outcomes, this mixed effects model object is merely used to
extract the required data (response vector, design matrices for fixed and random effects),
and starting values for the parameters and random effects. The definition of 
\code{densLong} and the call to \code{jointModelBayes()} take the form:
\begin{Schunk}
\begin{Sinput}
R> dLongBin <- function (y, eta.y, scale, log = FALSE, data) {
+      dbinom(x = y, size = 1, prob = plogis(eta.y), log = log)
+  }
R> jointFit.pbc3 <- jointModelBayes(lmeFit.pbc2, coxFit.pbc1, timeVar = "year", 
+                                   densLong = dLongBin)
R> summary(jointFit.pbc3)
\end{Sinput}
\end{Schunk}
\begin{Schunk}
\begin{Soutput}
. . .

Variance Components:
             StdDev    Corr
(Intercept)  7.1727  (Intr)
year         1.2210  0.4762
                           

Coefficients:
Longitudinal Process
              Value Std.Err Std.Dev    2.5%   97.5%      P
(Intercept) -1.7673  0.0913  0.5231 -2.8298 -0.8348 <0.001
year         0.9617  0.0187  0.1364  0.6920  1.2318 <0.001



Event Process
                     Value Std.Err  Std.Dev    2.5%    97.5%      P
drugD-penicil      -0.5485  0.0867   0.9381 -2.4007   1.2577  0.594
age                 0.0418  0.0026   0.0126  0.0177   0.0674 <0.001
drugD-penicil:age   0.0054  0.0016   0.0177 -0.0285   0.0401  0.772
Assoct              0.2188  0.0032   0.0320  0.1645   0.2869 <0.001
Bs.gammas1         -5.6968  0.1289   0.7261 -7.1613  -4.2927 <0.001

. . .
\end{Soutput}
\end{Schunk}
As we have already seen, the default parameterization posits that the subject-specific 
linear predictor $\eta_i(t)$ from the mixed model is included as a time-varying covariate 
in the relative risk model. This means that, in this case, the estimate of the 
association parameter $\alpha =$ 0.2 denotes the 
log hazard ratio for a unit increase in the log odds of having serum bilirubin above 
1.8 mg/dL. The flexibility that the user has in defining her own density function for the 
longitudinal outcome is evident, for example, we can easily fit a mixed effects probit 
regression instead of using the logit link by defining \code{densLong} as
\begin{Schunk}
\begin{Sinput}
R> dLongBin <- function (y, eta.y, scale, log = FALSE, data) {
+      dbinom(x = y, size = 1, prob = pnorm(eta.y), log = log)
+  }
\end{Sinput}
\end{Schunk}

As a final example, we illustrate how \code{densLong} can be utilized to fit joint models 
with censored longitudinal data (detection limit problem) by making use of extra variables
in the data frame containing the longitudinal information. Similarly to the previous 
example, the biomarkers collected in PBC study were not subject to detection limits, and 
therefore we again artificially create a censored version of serum bilirubin with values 
below the threshold value of 0.8 mg/dL set equal to the detection limit of 0.8 mg/dL. The 
code creating the censored longitudinal response vector is:
\begin{Schunk}
\begin{Sinput}
R> pbc2$CensInd <- as.numeric(pbc2$serBilir <= 0.8)
R> pbc2$serBilir2 <- pbc2$serBilir
R> pbc2$serBilir2[pbc2$serBilir2 <= 0.8] <- 0.8
\end{Sinput}
\end{Schunk}
In addition to the censored version of serum bilirubin we have also included in the data 
frame \code{pbc2} the censoring indicator \code{CensInd}. We again assume a normal error
distribution for the logarithm of serum bilirubin but in the definition of the 
corresponding density function we need to account for censoring, that is for observations 
above the detection limit we use the density function whereas for observations below this
limit we use the cumulative distribution function. The definition of the censored density 
becomes:
\begin{Schunk}
\begin{Sinput}
R> censdLong <- function (y, eta.y, scale, log = FALSE, data) {
+      log.f <- dnorm(x = y, mean = eta.y, sd = scale, log = TRUE)
+      log.F <- pnorm(q = y, mean = eta.y, sd = scale, log.p = TRUE)
+      ind <- data$CensInd
+      log.dens <- (1 - ind) * log.f + ind * log.F
+      if (log) log.dens else exp(log.dens)
+  }
\end{Sinput}
\end{Schunk}
Note that the censoring indicator is extracted from the \code{data} argument of 
\code{censdLong()}. Again in order to fit the joint model, we first need to fit the linear
mixed model for the censored response variable \code{serBilir2} and following supply this
object and the \code{censdLong()} to \code{jointModelBayes()}, i.e.,
\begin{Schunk}
\begin{Sinput}
R> lmeFit.pbc3 <- lme(log(serBilir2) ~ ns(year, 2), data = pbc2,
+                     random = ~ ns(year, 2) | id)
R> jointFit.pbc4 <- jointModelBayes(lmeFit.pbc3, coxFit.pbc1, timeVar = "year",
+                                    densLong = censdLong)
R> summary(jointFit.pbc4)
\end{Sinput}
\end{Schunk}
\begin{Schunk}
\begin{Soutput}
. . .

Variance Components:
              StdDev    Corr        
(Intercept)   1.2245  (Intr)  n(,2)1
ns(year, 2)1  2.5692  0.2089        
ns(year, 2)2  2.3330  0.2112  0.4860
Residual      0.3289                

Coefficients:
Longitudinal Process
              Value Std.Err Std.Dev   2.5%  97.5%      P
(Intercept)  0.3564  0.0163  0.0691 0.2263 0.5048 <0.001
ns(year, 2)1 2.4286  0.0577  0.2132 2.0175 2.8487 <0.001
ns(year, 2)2 2.1982  0.0751  0.3169 1.5572 2.7724 <0.001

Event Process
                     Value Std.Err  Std.Dev    2.5%    97.5%      P
drugD-penicil      -0.7645  0.0673   0.7172 -2.1080   0.6599  0.284
age                 0.0374  0.0014   0.0092  0.0186   0.0543 <0.001
drugD-penicil:age   0.0137  0.0013   0.0139 -0.0141   0.0409  0.321
Assoct              1.3871  0.0058   0.0964  1.1985   1.5816 <0.001
Bs.gammas1         -6.6386  0.0870   0.5591 -7.6702  -5.4673 <0.001

. . .
\end{Soutput}
\end{Schunk}
We observe that the estimate of the association parameter $\alpha$ is relatively close to 
the estimate obtained in model \code{jointFit.pbc1} that was based on the original 
(uncensored) version of serum bilirubin.


\subsection[Association structures]{Association structures} \label{Sec:JMbayesBasicUse-Ass}
The joint models we fitted in Sections~\ref{Sec:JMbayesBasicUse-Std} and 
\ref{Sec:JMbayesBasicUse-Ext} assumed that the hazard for an event at any time $t$ is 
associated with the current underlying value of the biomarker at the same time point, 
denoted as $\eta_i(t)$, and the strength of this association is measured by parameter 
$\alpha$. Even though under this formulation parameter $\alpha$ enjoys a clear 
interpretation, it is not realistic to expect that it will always be the most appropriate 
in expressing the correct relationship between the two processes. In general, there could 
be other characteristics of the subjects' longitudinal profiles that are more strongly 
predictive for the risk of an event; for example, the rate of increase/decrease of the 
biomarker's levels or a suitable summary of the whole longitudinal trajectory, among 
others. In this section we illustrate how such association structures could be postulated 
and fitted with \code{jointModelBayes()}.

We start with the parameterization proposed by \citet{ye.et.al:08b}, \citet{brown:09} and 
\citet{rizopoulos:12} that posits that the risk depends on both the current true value
of the trajectory and its slope at time $t$. More specifically, the relative risk survival
submodel takes the form,
\begin{equation}
h_i(t) = h_0(t) \exp \bigl \{ \bfgamma^\top \bw_i + \alpha_1 \eta_i(t)  + 
\alpha_2 \eta_i'(t) 
\bigr \}, \label{Eq:Param-TDslopes}
\end{equation}
where $\eta_i'(t) = d\{\bx_i^\top (t) \bfbeta + \bz_i^\top (t) \bb_i\} / dt$. The 
interpretation of parameter $\alpha_1$ remains the same as in the standard 
parameterization. Parameter $\alpha_2$ measures the association between the 
slope of the true longitudinal trajectory at time $t$ and the risk for an event at the 
same time point, provided that $\eta_i(t)$ remains constant. To fit the joint model with 
the extra slope term in the relative risk submodel we need to specify the \code{param} and
\code{extraForm} arguments of \code{jointModelBayes()}. The first one is a character
string with options \code{"td-value"} (default) that denotes that only the current 
value term $\eta_i(t)$ is included, \code{"td-extra"} which means that only the extra, 
user-defined, term is included, and \code{"td-both"} which means that both $\eta_i(t)$ and
the user-defined terms are included. The exact definition of the extra term is provided 
via the argument \code{extraForm} which is a list with four components, namely
\begin{itemize}
\item[*] \code{"fixed"} an \proglang{R} formula specifying the fixed-effects part of the 
extra term,

\item[*] \code{"random"} an \proglang{R} formula specifying the random-effects part of the 
extra term,

\item[*] \code{"indFixed"} an integer vector denoting which of the fixed effects of the 
original mixed model are encountered in the definition of the extra term, and

\item[*] \code{"indRandom"} an integer vector denoting which of the random effects of the 
original mixed model are encountered in the definition of the extra term.
\end{itemize}
For example, to include the slope term $\eta_i'(t)$ under the linear mixed model 
\code{lmeFit.pbc1}, this list takes the form:
\begin{Schunk}
\begin{Sinput}
R> dForm <- list(fixed = ~ 0 + dns(year, 2), random = ~ 0 + dns(year, 2), 
+                indFixed = 2:3, indRandom = 2:3)
\end{Sinput}
\end{Schunk}
Function \code{dns()} computes numerically (with a central difference approximation) the
derivative of a natural cubic spline as calculated by function \code{ns()} (there is also 
a similar function \code{dbs()} that computes numerically the derivative of a cubic spline
as calculated by function \code{bs()}). The corresponding joint model is fitted with the 
code:
\begin{Schunk}
\begin{Sinput}
R> jointFit.pbc12 <- update(jointFit.pbc1, param = "td-both", 
+                           extraForm = dForm)
R> summary(jointFit.pbc12)
\end{Sinput}
\end{Schunk}
\begin{Schunk}
\begin{Soutput}
. . .

Event Process
                     Value Std.Err  Std.Dev    2.5%    97.5%      P
drugD-penicil      -0.1915  0.0616   0.7861 -1.6745   1.4312  0.801
age                 0.0441  0.0014   0.0102  0.0236   0.0652 <0.001
drugD-penicil:age   0.0031  0.0012   0.0150 -0.0271   0.0318  0.817
Assoct              1.3282  0.0060   0.1059  1.1281   1.5476 <0.001
AssoctE             2.6043  0.0505   0.5898  1.4132   3.6976 <0.001
Bs.gammas1         -8.0084  0.0804   0.6517 -9.2579  -6.7369 <0.001

. . .
\end{Soutput}
\end{Schunk}
We observe that both the current level and the current slope of the longitudinal profile 
are strongly associated with the risk for the composite event. For patients with the same 
treatment and age at baseline, and who have the same underlying level of serum bilirubin 
at time $t$, if serum bilirubin has increased by 50\% within a year then the 
corresponding hazard ratio is 2.9 (95\% CI: 1.8; 
4.5).

A common characteristic of the two parameterizations we have seen so far is that the risk 
for an event at any time $t$ is assumed to be associated with features of the longitudinal 
trajectory at the same time point (i.e., current value $\eta_i(t)$ and current slope 
$\eta_i'(t)$). However, this assumption may not always be appropriate, and we may benefit 
from allowing the risk to depend on a more elaborate function of the history of the 
time-varying covariate \citep{sylvestre.abrahamowicz:09}. In the context of joint models, 
one option to account for the cumulative effect of the longitudinal outcome is to include 
in the linear predictor of the relative risk submodel the integral of the longitudinal 
trajectory from baseline up to time $t$ \citep{brown:09, rizopoulos:12}. More 
specifically, the survival submodel takes the form
\[
h_i(t) = h_0(t) \exp \biggl \{ \bfgamma^\top \bw_i + \alpha \int_0^t \eta_i(s)
ds \biggr \},
\]
where for any particular time point $t$, $\alpha$ measures the strength of the association
between the risk for an event at time point $t$ and the area under the longitudinal 
trajectory up to the same time $t$, with the area under the longitudinal trajectory taken 
as a summary of the whole marker history $\mathcal H_i(t) = \{ m_i(s), 0 \leq s < t \}$. 
To fit a joint model with this term in the linear predictor of the survival submodel, we
need first again to appropriately define the formulas that calculate its fixed-effects and
random-effects parts. Similarly to including the slope term, the list with these formulas
takes the form
\begin{Schunk}
\begin{Sinput}
R> iForm <- list(fixed = ~ 0 + year + ins(year, 2), 
+                random = ~ 0 + year + ins(year, 2), 
+                indFixed = 1:3, indRandom = 1:3)
\end{Sinput}
\end{Schunk}
where function \code{ins()} calculates numerically (using the Gauss-Kronrod rule) the 
integral of function \code{ns()}. The corresponding joint model is fitted by supplying 
this list in the \code{extraForm} argument and by also setting in argument \code{param} 
that we only want to include the extra term in the linear predictor of the survival 
submodel:
\begin{Schunk}
\begin{Sinput}
R> jointFit.pbc13 <- update(jointFit.pbc1, param = "td-extra", 
+                           extraForm = iForm)
R> summary(jointFit.pbc13)
\end{Sinput}
\end{Schunk}
\begin{Schunk}
\begin{Soutput}
. . .

Event Process
                     Value Std.Err  Std.Dev     2.5%    97.5%      P
drugD-penicil      -0.7543  0.0781   0.7550  -2.3085   0.7369  0.306
age                 0.0365  0.0015   0.0107   0.0144   0.0564 <0.001
drugD-penicil:age   0.0096  0.0015   0.0144  -0.0185   0.0382  0.505
AssoctE             0.2272  0.0012   0.0200   0.1884   0.2650 <0.001
Bs.gammas1         -4.5973  0.0729   0.6197  -5.7501  -3.3098 <0.001

. . .
\end{Soutput}
\end{Schunk}
To explicitly denote that in the relative risk submodel we only want to include the 
user-defined integral term, we have set argument \code{param} to \code{"td-extra"}. 
Similarly to the previous results we observe that the area under the longitudinal 
profile of log serum bilirubin is strongly associated with the risk for an event, with a 
unit increase corresponding to a 1.3-fold 
(95\% CI: 1.2; 1.3) increase of the risk.

The final type of association structure we consider assumes that only the random effects 
are shared between the two processes, namely
\begin{equation}
h_i(t) = h_0(t) \exp (\bfgamma^\top \bw_i + \bfalpha^\top \bb_i), 
\label{Eq:Param-RE}
\end{equation}
or potentially the corresponding fixed effects may also be included, i.e.,
\begin{equation}
h_i(t) = h_0(t) \exp \{ \bfgamma^\top \bw_i + \bfalpha^\top (\bfbeta_b + \bb_i) \},
\label{Eq:Param-fixedRE}
\end{equation}
with $\bfbeta_b$ denoting the fixed effects that correspond to the random effects. This 
type of parameterization is more meaningful when a simple random-intercepts and 
random-slopes structure is assumed for the longitudinal submodel, in which case the random
effects express subject-specific deviations from the average intercept and average slope. 
Under this setting this parameterization postulates that patients who have a lower/higher 
level for the longitudinal outcome at baseline (i.e., intercept) or who show a steeper 
increase/decrease in their longitudinal trajectories (i.e., slope) are more likely to 
experience the event. In that respect, this formulation shares also similarities with the 
time-dependent slopes formulation \eqref{Eq:Param-TDslopes}. A joint model with a relative
risk model of the form \eqref{Eq:Param-RE} can be fitted by setting argument 
\code{param} to \code{"shared-RE"} in the call to \code{jointModelBayes()}, whereas 
formulation \eqref{Eq:Param-fixedRE} is specified by setting \code{param} to 
\code{"shared-betasRE"}; for example, for the PBC dataset a joint model with this 
parameterization is fitted with the code
\begin{Schunk}
\begin{Sinput}
R> jointFit.pbc14 <- update(jointFit.pbc1, param = "shared-betasRE", 
+                           n.iter = 50000)
R> summary(jointFit.pbc14)
\end{Sinput}
\end{Schunk}
\begin{Schunk}
\begin{Soutput}
. . .

Event Process
                       Value Std.Err Std.Dev     2.5%    97.5%      P
drugD-penicil        -0.0336  0.1143  1.0687  -2.0559   2.0417  0.952
age                   0.0503  0.0033  0.0166   0.0222   0.0850 <0.001
drugD-penicil:age    -0.0006  0.0023  0.0206  -0.0412   0.0378  0.994
Assoct:(Intercept)    1.2932  0.0184  0.1804   0.9525   1.6719 <0.001
Assoct:ns(year, 2)1   0.5415  0.0062  0.0741   0.3969   0.6849 <0.001
Assoct:ns(year, 2)2   0.2902  0.0194  0.1228   0.0735   0.5647  0.005
Bs.gammas1          -12.0474  0.3221  1.5382 -15.2916  -9.5855 <0.001

. . .
\end{Soutput}
\end{Schunk}
The results suggest that both the baseline levels of the underlying log serum bilirubin 
(i.e., parameter \code{Assoct:(Intercept)}) as well as the longitudinal evolution of the 
marker (i.e., parameters \code{Assoct:ns(year, 2)1} and \code{Assoct:ns(year, 2)2}) are 
strongly related to the hazard of the composite event.


\subsection[Tranformation functions]{Transformation functions} 
\label{Sec:JMbayesBasicUse-Transf}
The previous section illustrated several options for the definition of function $f(\cdot)$
in \eqref{Eq:Surv-RR} for studying which features of the longitudinal process are 
associated with the event of interest. Yet another set of options for function $f(\cdot)$ 
would be to consider adding interaction or nonlinear terms for the components 
of the longitudinal outcome that are included in the linear predictor of the relative risk
model. Such options are provided in \code{jointModelBayes()} by suitably specifying 
argument \code{transFun}. This should be a function (or a list of two functions) with 
arguments \code{x} denoting the term from the longitudinal model, and \code{data} a data 
frame that contains other variables that potentially should be included in the calculation. 
When a single function is provided, then this function is applied to the current value term 
$\eta_i(t)$ and potentially also to the extra term provided by the user if \code{param} 
was set to \code{"td-both"}. If a list is provided, then this should be a named list with
components \code{"value"} and \code{"extra"} providing separate functions for the current
value and the user-defined terms, respectively. We illustrate how these transformation 
functions can be used in practice by extending model \code{jointFit.pbc12}, which included
the current value term $\eta_i(t)$ and the current slope term $\eta_i'(t)$, by including
the quadratic effect of $\eta_i(t)$ and the interaction of $\eta_i'(t)$ with the 
randomized treatment, i.e.,
\begin{eqnarray*}
h_i(t) & = & h_0(t) \exp \bigl [ \gamma_1 \mbox{\tt D-penicil}_i +
\gamma_2 \mbox{\tt Age}_i + \gamma_3 (\mbox{\tt D-penicil}_i \times \mbox{\tt Age}_i)\\
&  & \hspace{1cm} + \, \alpha_1 \eta_i(t) + \alpha_2 \{\eta_i(t)\}^2 +
\alpha_3 \eta_i'(t) + \alpha_4 \{\eta_i'(t) \times \mbox{\tt D-penicil}_i\} \bigr ].
\end{eqnarray*}
To fit the corresponding joint model we first define the two transformation functions as:
\begin{Schunk}
\begin{Sinput}
R> tf1 <- function (x, data) {
+      cbind(x, "^2" = x*x)
+  }
R> tf2 <- function (x, data) {
+      cbind(x, "D-penicil" = x * (data$drug == 'D-penicil'))
+  }
\end{Sinput}
\end{Schunk}
Following we update the call to \code{jointFit.pbc12} and supply the list of the two
functions in argument \code{transFun},
\begin{Schunk}
\begin{Sinput}
R> jointFit.pbc15 <- update(jointFit.pbc12, 
+                           transFun = list("value" = tf1, "extra" = tf2))
R> summary(jointFit.pbc15)
\end{Sinput}
\end{Schunk}
\begin{Schunk}
\begin{Soutput}
. . .

Event Process
                     Value Std.Err  Std.Dev    2.5%    97.5%      P
drugD-penicil      -0.8048  0.1224   0.9507 -2.7879   0.9564  0.395
age                 0.0368  0.0039   0.0134  0.0098   0.0595  0.015
drugD-penicil:age   0.0125  0.0016   0.0182 -0.0199   0.0485  0.508
Assoct              0.7052  0.0301   0.3308  0.0790   1.3689  0.031
Assoct:^2           0.1839  0.0072   0.0936  0.0020   0.3693  0.044
AssoctE             2.6437  0.0620   0.7448  1.1891   4.1787 <0.001
AssoctE:D-penicil   0.4071  0.0925   1.0102 -1.5416   2.3909  0.677
Bs.gammas1         -7.3531  0.2437   0.8136 -8.8028  -5.7138 <0.001

. . .
\end{Soutput}
\end{Schunk}
There is some weak evidence that the effect of $\eta_i(t)$ could be nonlinear, but clearly
the association between $\eta_i'(t)$ and the hazard does not seem to be different between 
the two treatment groups. 


\subsection[Supporting functions]{Supporting functions} 
\label{Sec:JMbayesBasicUse-Methods}
Several supporting functions are available in the package that extract or calculate useful
statistics based on the fitted joint model. In particular, function 
\code{jointModelBayes()} return objects of class \code{"JMbayes"}, for which there are 
\proglang{S}3 methods defined for several of the standard generic functions in 
\proglang{R}. The most important are enlisted below:
\begin{description}
\item Functions \code{\textbf{coef()}} and \textbf{\code{fixef()}} extract the estimated 
coefficients for the two submodels from a fitted joint model. For the survival process 
both provide the same output, but for the longitudinal model, the former returns the 
subject-specific regression coefficients (i.e., the fixed effects plus their corresponding
random effects estimates), whereas the latter only returns the estimated fixed effects.

\item Function \textbf{\code{ranef()}} extracts the empirical Bayes estimates for the 
random effects for each subject. The function also extracts estimates for the dispersion 
matrix of the posterior of the random effects using argument \code{postVar}.

\item Function \code{vcov()} extracts the estimated variance-covariance matrix of the 
parameters from the MCMC sample.

\item Functions \code{fitted()} and \code{residuals()} compute several kind of fitted 
values and residuals, respectively, for the two outcomes.  For the longitudinal outcome 
the \code{fitted()} method computes the marginal $\bX \hat \bfbeta$ and 
subject-specific $\bX \hat \bfbeta + \bZ \hat \bb_i$ fitted values, where $\hat \bfbeta$ 
and $\hat \bb_i$ denote the posterior means of the fixed and random effects, whereas for 
the survival outcome it computes the cumulative hazard function for every subject and 
every time point a longitudinal measurement was collected. Analogously, method 
\code{residuals()} calculates the marginal and subject-specific residuals for the 
longitudinal outcome, and the martingale residuals for the survival outcome. 

\item Function \code{anova()} can be used to compare joint models on the basis of 
the DIC, pD and LPML values.

\item Function \code{plot()} produces diagnostic plots for the MCMC sample, including 
trace plots, auto-correlation plots and kernel density estimation plots.

\item Function \code{predict()} produces subject-specific and 
marginal predictions for the longitudinal outcome, while function \code{survfitJM()} 
produces subject-specific predictions for the event time outcome, along with associated 
confidence or prediction confidence intervals. The use of these functions is explained 
in detail and illustrated in Section~\ref{Sec:DynPred}.

\item Function \code{logLik()} calculates the log-likelihood value for the posterior means
of the parameters and the random effects, and can be also used to obtain the marginal 
log-likelihood (integrated over the parameters and random effects) using the Laplace 
approximation.

\item Function \code{xtable()} returns the \LaTeX{} code to produce the table of 
posterior means, posterior standard deviations, and 95\% credibility intervals from a 
fitted joint model. This is a method for the generic function \code{xtable()} from 
package \pkg{xtable} \citep{xtable}.
\end{description}


\section[Dynamic predictions]{Dynamic predictions} \label{Sec:DynPred}
\subsection[Definitions and estimation]{Definitions and estimation}\label{Sec:DynPred-Est}
In recent years there has been increasing interest in medical research towards personalized
medicine. In particular, physicians would like to tailor decision making on the 
characteristics of individuals patients with aim to optimize medical care. In the same 
sense, patients who are informed about their individual health risk often decide to adjust
their lifestyles to mitigate it. In this context it is often of interest to utilize 
results from tests performed on patients on a regular basis to derive medically-relevant 
summary measures, such as survival probabilities. Joint models constitute a valuable tool 
that can be used to derive such probabilities and also provide predictions for future 
biomarker levels. More specifically, under the Bayesian specification of the joint model, 
presented in Section~\ref{Sec:Theory}, we can derive subject-specific predictions for 
either the survival or longitudinal outcomes \citep{yu.et.al:08, rizopoulos:11, 
rizopoulos:12, taylor.et.al:13}. To put it more formally, based on a joint model fitted in
a sample $\mathcal D_n = \{T_i, \delta_i, \by_i; i = 1, \ldots, n\}$ from the target 
population, we are interested in deriving predictions for a new subject $j$ from the same 
population that has provided a set of longitudinal measurements $\mathcal Y_j(t) = \{ 
y_j(t_{jl}); 0 \leq t_{jl} \leq t, l = 1, \ldots, n_j \}$, and has a vector of baseline 
covariates $\bw_j$. The fact that biomarker measurements have been recorded up to $t$, 
implies that subject $j$ was event-free up to this time point, and therefore it is more 
relevant to focus on conditional subject-specific predictions, given survival up to $t$. 
In particular, for any time $u > t$ we are interested in the probability that subject $j$ 
will survive at least up to $u$, i.e.,
\[
\pi_j(u \mid t) = \Pr ( T_j^* \geq u \mid T_j^* > t, \mathcal Y_j(t), \bw_j, \mathcal D_n).
\]
Similarly, for the longitudinal outcome we are interested in the predicted longitudinal 
response at $u$, i.e.,
\[
\omega_j(u \mid t) = E \bigl \{ y_j(u) \mid T_j^* > t, \mathcal Y_j(t), 
\mathcal D_n \bigr \}.
\]
The time-dynamic nature of both $\pi_j(u \mid t)$ and $\omega_j(u \mid t)$ is evident from
the fact that when new information is recorded for subject $j$ at time $t' > t$, we can 
update these predictions to obtain $\pi_j(u \mid t')$ and $\omega_j(u \mid t')$, and 
therefore proceed in a time-dynamic manner.

Under the joint modeling framework of Section~\ref{Sec:Theory}, estimation of either 
$\pi_j(u \mid t)$ or $\omega_j(u \mid t)$ is based on the corresponding posterior 
predictive distributions, namely
\[
\pi_j(u \mid t) = \int \Pr(T_j^* \geq u \mid T_j^* > t, \mathcal Y_j(t), \bftheta) \, 
p(\bftheta \mid \mathcal D_n) \, d\bftheta,
\]
for the survival outcome, and analogously
\[
\omega_j(u \mid t) = \int E \bigl \{ y_j(u) \mid T_j^* > t, \mathcal Y_j(t), 
\bftheta \bigr \} \, p(\bftheta \mid \mathcal D_n) \, d\bftheta,
\]
for the longitudinal one. The calculation of the first part of each integrand takes full 
advantage of the conditional independence assumptions \eqref{Eq:CondInd-I} and 
\eqref{Eq:CondInd-II}. In particular, we observe that the first term of the integrand of
$\pi_j(u \mid t)$ can be rewritten by noting that:
\begin{eqnarray*}
\nonumber \Pr(T_j^* \geq u \mid T_j^* > t, \mathcal Y_j(t), \bftheta) & = & \int 
\Pr(T_j^* \geq u \mid T_j^* > t, \bb_j, \bftheta) \, p(\bb_j \mid T_j^* > t, \mathcal 
Y_j(t), \bftheta) \, d\bb_j\\
& = & \int \frac{S_j \bigl \{ u \mid \mathcal H_j(u, \bb_j), \bftheta \bigr \}}{S_j 
\bigl \{ t \mid \mathcal H_j(t, \bb_j), \bftheta \bigr \} } \, p(\bb_j \mid T_j^* > t, 
\mathcal Y_j(t), \bftheta) \, d\bb_j,
\end{eqnarray*}
whereas for $\omega_j(u \mid t)$ we similarly have:
\begin{eqnarray*}
\nonumber E \bigl \{ y_j(u) \mid T_j^* > t, \mathcal Y_j(t), \bftheta \bigr \} & = & 
\int E \bigl \{ y_j(u) \mid \bb_j, \bftheta \} \,
p(\bb_j \mid T_j^* > t, \mathcal Y_j(t), \bftheta) \, d\bb_j\\
& = &  \bx_j^\top(u) \bfbeta + \bz_j^\top(u) \bar \bb_j^{(t)},
\end{eqnarray*}
with
\[
\bar \bb_j^{(t)} = \int \bb_j \, p(\bb_j \mid T_j^* > t, \mathcal Y_j(t), 
\bftheta) \, d\bb_j.
\]
Combining these equations with the MCMC sample from the posterior distribution of the 
parameters for the original data $\mathcal D_n$, we can devise a simple simulation scheme 
to obtain Monte Carlo estimates of $\pi_j(u \mid t)$ and $\omega_j(u \mid t)$. More 
details can be found in \citet{yu.et.al:08}, \citet{rizopoulos:11, rizopoulos:12}, and 
\citet{taylor.et.al:13}.

\begin{figure}
\centering
\includegraphics[width = 0.9\textwidth]{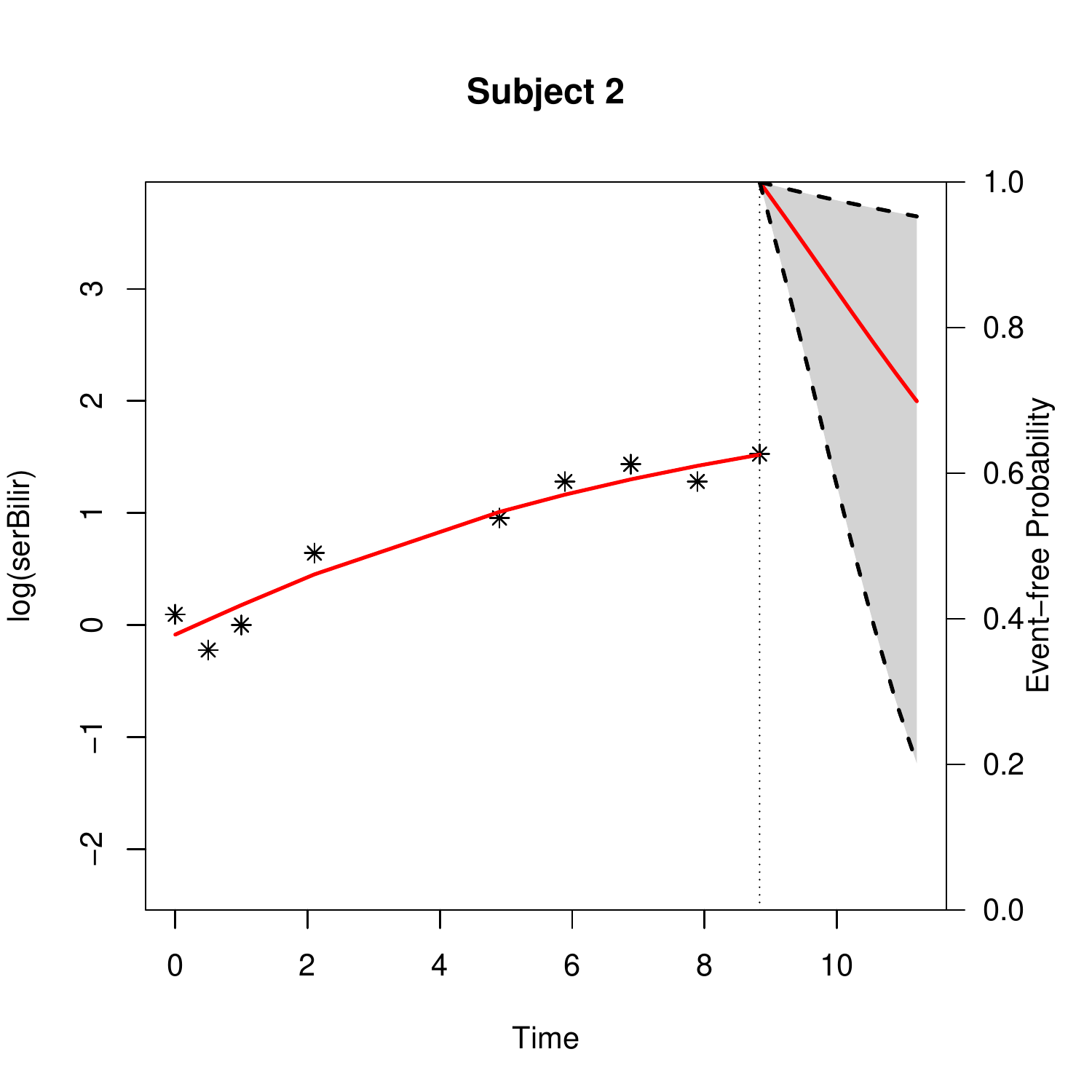}
\caption{Estimated conditional survival probabilities for Patient 2 from
the PBC dataset.} \label{Fig:PredSurv}
\end{figure}

In package \pkg{JMbayes} these subject-specific predictions for the survival and 
longitudinal outcomes can be calculated using functions \code{survfitJM()} and 
\code{predict()}, respectively. As an illustration we show how these functions can be 
utilized to derive predictions for Patient~2 from the PBC dataset using joint model 
\code{jointFit.pbc15}. We first extract the data of this patient in a separate data frame
\begin{Schunk}
\begin{Sinput}
R> ND <- pbc2[pbc2$id == 2, ]
\end{Sinput}
\end{Schunk}
Function \code{survfitJM()} has two required arguments, the joint model object based on 
which predictions will be calculated, and the data frame with the available longitudinal 
data and baseline information. For Patient~2 estimates of $\pi_j(u \mid t)$ are calculated
with the code:
\begin{Schunk}
\begin{Sinput}
R> sfit.pbc15 <- survfitJM(jointFit.pbc15, newdata = ND)
R> sfit.pbc15
\end{Sinput}
\begin{Soutput}
Prediction of Conditional Probabilities for Event
	based on 200 Monte Carlo samples

$`2`
    times   Mean Median  Lower  Upper
1  8.8325 1.0000 1.0000 1.0000 1.0000
1  8.9232 0.9888 0.9908 0.9696 0.9978
2  9.2496 0.9475 0.9575 0.8570 0.9904
3  9.5759 0.9055 0.9237 0.7404 0.9834
4  9.9022 0.8631 0.8898 0.6177 0.9767
5 10.2286 0.8208 0.8550 0.5026 0.9703
6 10.5549 0.7791 0.8214 0.3925 0.9642
7 10.8812 0.7384 0.7864 0.2900 0.9583
8 11.2076 0.6989 0.7482 0.2016 0.9527
\end{Soutput}
\end{Schunk}
By default \code{survfitJM()} assumes that the patient was event-free up to the time point
of the last longitudinal measurement (if the patient was event-free up to a later time 
point, this can be specified using argument \code{last.time}). In addition, by default 
\code{survfitJM()} estimates of $\pi_j(u \mid t)$ using 200 Monte Carlo samples 
(controlled with argument \code{M}) for the time points $\{u: u > t_\ell, \ell = 1, 
\ldots, 35\}$, with $\{t_\ell, \ell = 1, \ldots, 35\}$ calculated as \code{seq(min(Time), 
quantile(Time, 0.9) + 0.01, length.out = 35)} with \code{Time} denoting the observed event
times variable. The user may override these default time points and specify her own using
argument \code{survTimes}. In the output of \code{survfitJM()} we obtain as estimates of 
$\pi_j(u \mid t)$ the mean and median over the Monte Carlo samples along with the 
95\% pointwise confidence intervals. If only point estimates are of interest, 
\code{survfitJM()} provides the option (by setting argument \code{simulate} to 
\code{FALSE}) to use a first order estimator of $\pi_j(u \mid t)$ calculated as:
\[
\tilde \pi_j(u \mid t) = \frac{S_j \bigl \{ u \mid \mathcal H_j(u, \hat \bb_j), 
\hat \bftheta \bigr \}}{S_j \bigl \{ t \mid \mathcal H_j(t, \hat \bb_j), \hat \bftheta 
\bigr \} },
\]
where $\hat \bftheta$ denotes here the posterior means of the model parameters, and 
$\hat \bb_j$ the mode of the posterior density $p(\bb_j \mid T_j^* > t, 
\mathcal Y_j(t), \hat \bftheta)$ with respect to $\bb_j$. The corresponding \code{plot()} 
method for objects created by \code{survfitJM()} produces the figure of estimated 
conditional survival probabilities; for Patient~2 this is depicted in 
Figure~\ref{Fig:PredSurv}. By setting logical argument \code{include.y} to \code{TRUE}, 
the fitted longitudinal profile is also included in the plot, i.e.,
\begin{Schunk}
\begin{Sinput}
R> plot(sfit.pbc15, estimator = "mean", include.y = TRUE,
+       conf.int = TRUE, fill.area = TRUE, col.area = "lightgrey")
\end{Sinput}
\end{Schunk}
Argument \code{estimator} specifies whether the \code{"mean"} or the \code{"median"} over
the 200 Monte Carlo samples should be used as an estimate of $\pi_j(u \mid t)$, and in 
addition arguments \code{conf.int}, \code{fill.area} and \code{col.area} control the 
appearance of the 95\% confidence intervals. In a similar manner, predictions for the 
longitudinal outcome are calculated by the \code{predict()} function. For example, 
predictions of future log serum bilirubin levels for Patient~2 are produced with the code:
\begin{Schunk}
\begin{Sinput}
R> Ps.pbc15 <- predict(jointFit.pbc15, newdata = ND, type = "Subject", 
+                      interval = "confidence", return = TRUE)
\end{Sinput}
\end{Schunk}
Argument \code{type} specifies if subject-specific or marginal predictions are to be 
computed\footnote{by marginal predictions we refer to $\bx_i^\top(t) \hat \bfbeta$, 
whereas by subject-specific to $\bx_i^\top(t) \hat \bfbeta + \bz_i^\top(t) \hat \bb_i$.}, 
argument \code{interval} specifies the type of interval to compute (i.e., confidence or 
prediction), and by setting argument \code{return} to \code{TRUE}, \code{predict()} 
returns the data frame supplied in the required argument \code{newdata} having as extra 
columns the corresponding predictions and the limits of the confidence/prediction 
interval. This option facilitates plotting these predictions by a simple call to 
\code{xyplot()}, i.e.,
\begin{Schunk}
\begin{Sinput}
R> last.time <- with(Ps.pbc15, year[!is.na(low)][1])
R> xyplot(pred + low + upp ~ year, data = Ps.pbc15, type = "l", lty = c(1,2,2), 
+         col = c(2,1,1), abline = list(v = last.time, lty = 3), 
+         xlab = "Time (years)", ylab = "Predicted log(serum bilirubin)")
\end{Sinput}
\end{Schunk}
\begin{figure}
\centering
\includegraphics[width = 0.9\textwidth]{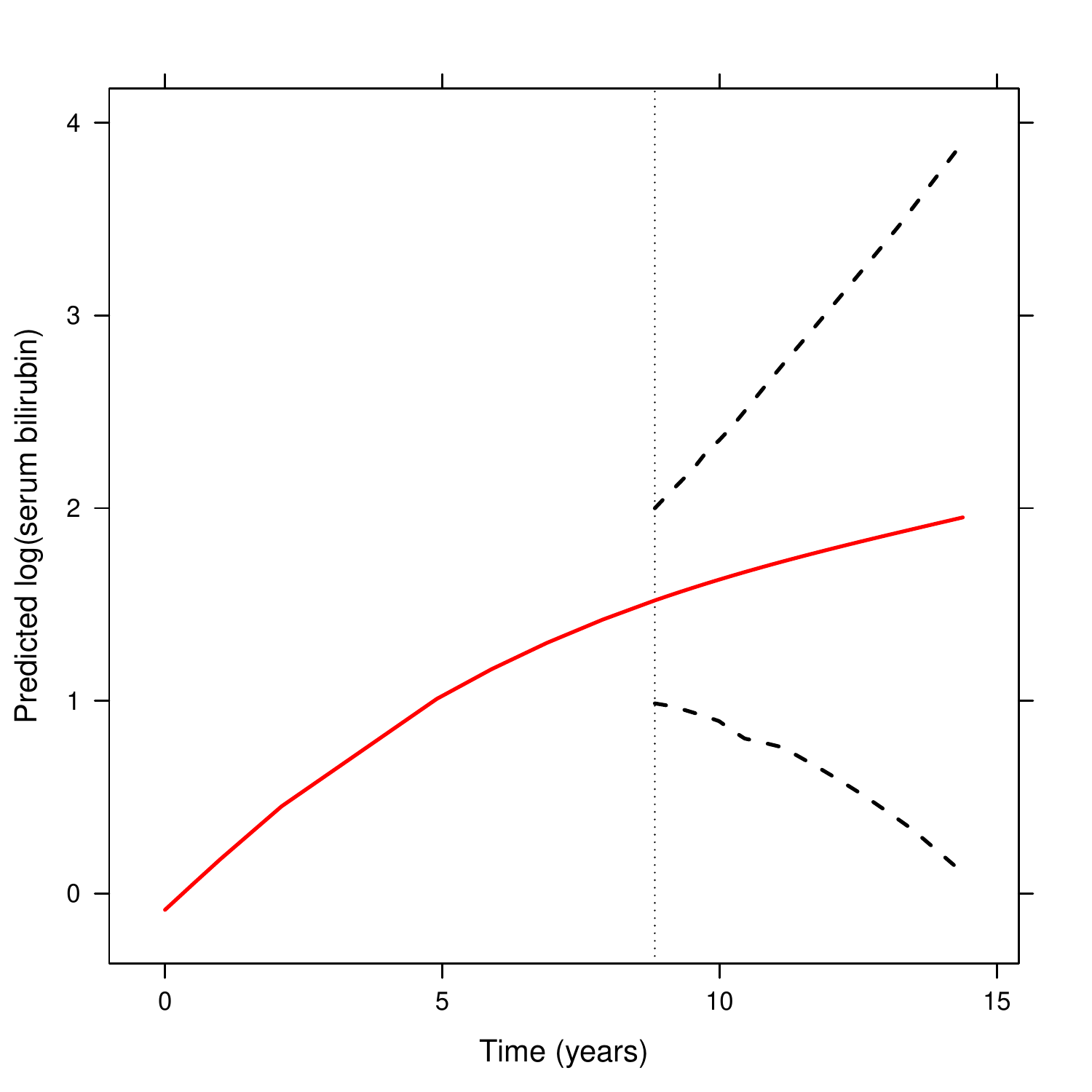}
\caption{Predicted longitudinal trajectory (with a 95\% pointwise confidence interval)
for Patient~2 from the PBC dataset. The dotted line denotes the last time point Patient~2
was still event-free.} \label{Fig:PredLong}
\end{figure}
The first line of the code extracts from the data frame \code{Ps.pbc15} the last time 
point at which Patient~2 was still alive, which is passed in the \code{panel} function
that produces Figure~\ref{Fig:PredLong}.


\subsubsection[Web interface using shiny]{Web interface using \pkg{shiny}}
To facilitate the use of package \pkg{JMbayes} for deriving individualized predictions, 
a web interface has been written using using package \pkg{shiny} \citep{shiny}. This is 
available in the demo folder of the package and can be invoked with the code (assuming 
that \pkg{JMbayes} has been installed in the default library):
\begin{Schunk}
\begin{Sinput}
R> library("shiny")
R> runApp(file.path(.Library, "JMbayes/demo"))
\end{Sinput}
\end{Schunk}
With this interface users may load an \proglang{R} workspace with the fitted joint 
model(s), following load the data of the new subject, and subsequently obtain dynamic 
estimates of $\pi_j(u \mid t)$ and $\omega_j(u \mid t)$ (i.e., an estimate after each 
longitudinal measurement). Several additional options are provided to calculate 
predictions based on different joint models (if the \proglang{R} workspace contains more 
than one models), to obtain estimates at specific horizon times, and to extract the 
dataset with the estimated conditional survival probabilities.

\newpage
\subsection[Bayesian model averaging]{Bayesian model averaging}\label{Sec:DynPred-BMA}
Section~\ref{Sec:JMbayesBasicUse-Ass} demonstrated that there are several choices to link 
the longitudinal and event time outcomes. When faced with this problem, the common 
practice in prognostic modeling is to base predictions on a single model that has been 
selected based on an automatic algorithm, such as, backward, forward or stepwise 
selection, or on likelihood-based information criteria, such as, AIC, BIC, DIC and their 
variants. However, what is often neglected in this procedure is the issue of model 
uncertainty. For example, if we choose a model using any of these criteria, say DIC, we 
usually treat it as the true model, even if there could be more than one models with DIC 
values of similar magnitude. In addition, when it comes to using a model for deriving 
predictions, we implicitly make the assumption that this model is adequate for all future 
patients. This seldom will be true in clinical practice. In our setting, a joint model 
with a specific formulation of the association structure may produce more accurate 
predictions for subjects with specific longitudinal profiles, while other models with 
other association structures may produce better predictions for subjects whose profiles 
have other characteristics. Here we follow another approach and we explicitly take into 
account model uncertainty by combining predictions under different association structures 
using Bayesian model averaging (BMA) \citep{hoeting.et.al:99, rizopoulos.et.al:14}.

We focus here on dynamic BMA predictions of survival probabilities. BMA predictions for 
the longitudinal outcome can be produced with similar methodology. Following the 
definitions of Section~\ref{Sec:DynPred-Est}, we assume that we have available data 
$\mathcal D_n = \{T_i, \delta_i, \by_i; i = 1, \ldots, n\}$ based on which we fit $M_1, 
\ldots, M_K$ joint models with different association structures. Interest is in 
calculating predictions for a new subject $j$ from the same population who has provided a 
set of longitudinal measurements $\mathcal Y_j(t)$, and has a vector of baseline 
covariates $\bw_j$. We let $\mathcal D_j(t) = \{T_j^* > t, \mathcal Y_j(t), \bw_j\}$ 
denote the available data for this subject. The model-averaged probability of subject 
$j$ surviving time $u > t$, given her survival up to $t$ is given by the expression:
\begin{equation}
\Pr (T_j^* > u \mid \mathcal D_j(t), \mathcal D_n) = \sum \limits_{k = 1}^K 
\Pr (T_j^* > u \mid M_k, \mathcal D_j(t), \mathcal D_n) \, p(M_k \mid \mathcal D_j(t), 
\mathcal D_n). \label{Eq:BMA-SurvPred}
\end{equation}
The first term in the right-hand side of \eqref{Eq:BMA-SurvPred} denotes the 
model-specific survival probabilities, derived in Section~\ref{Sec:DynPred-Est}, and the 
second term denotes the posterior weights of each of the competing joint models. The 
unique characteristic of these weights is that they depend on the observed data of subject
$j$, in contrast to classic applications of BMA where the model weights depend only on 
$\mathcal D_n$ and are the same for all subjects. This means that, in our case, the model 
weights are both subject- and time-dependent, and therefore, for different subjects, and 
even for the same subject but at different times points, different models may have higher 
posterior probabilities \citep{rizopoulos.et.al:14}. Hence, this framework is capable of 
better tailoring predictions to each subject than standard prognostic models, because at 
any time point we base risk assessments on the models that are more probable to describe 
the association between the observed longitudinal trajectory of a subject and the risk for
an event.

For the calculation of the model weights we observe that these are written as 
\citep{rizopoulos.et.al:14}:
\[
p(M_k \mid \mathcal D_j(t), \mathcal D_n) = \frac{p(\mathcal D_j(t) \mid M_k) \, 
p(\mathcal D_n \mid M_k) \, p(M_k)}{\sum \limits_{\ell = 1}^K 
p(\mathcal D_j(t) \mid M_\ell) \, p(\mathcal D_n \mid M_\ell) \, p(M_\ell)},
\]
where
\[
p(\mathcal D_j(t) \mid M_k) = \int p(\mathcal D_j(t) \mid \bftheta_k) 
p(\bftheta_k \mid M_k) \, d\bftheta_k
\]
and $p(\mathcal D_n \mid M_k)$ is defined analogously. The likelihood part 
$p(\mathcal D_n \mid \bftheta_k)$ is based on \eqref{Eq:FullPost}, and similarly 
$p(\mathcal D_j(t) \mid \bftheta_k)$ equals
\[
p(\mathcal D_j(t) \mid \bftheta_k) = p(\mathcal Y_j(t) \mid \bb_j, \bftheta_k) \, 
S_j(t \mid \bb_j, \bftheta_k) \, p(\bb_j \mid \bftheta_k).
\]
Thus, the subject-specific information in the model weights at time $t$ comes from the 
available longitudinal measurements $\mathcal Y_j(t)$ but also from the fact that this 
subject has survived up to $t$. We should note that the new subject $j$ does not 
contribute any information about $\bftheta_k$ (i.e., we do not refit the models using the 
data of this subject), the information for the parameters only comes from the original
dataset in which the joint models have been fitted via the posterior distribution 
$p(\bftheta_k \mid \mathcal D_n, M_k)$. A priori we assume that all models are equally 
probable, i.e., $p(M_k) = 1/K$, for all $k = 1, \ldots, K$. Closed-form expressions for 
the marginal densities $p(\mathcal D_n \mid M_k)$ and $p(\mathcal D_j(t) \mid M_k)$ are 
obtained by means of Laplace approximations \citep{tierney.kadane:86} performed in 
two-steps, namely, first integrating out the random effects and then the parameters.

In package \pkg{JMbayes} BMA predictions for either the survival or longitudinal outcome 
can be calculated using function \code{bma.combine()}. This function accepts a series or a
list of objects returned by either \code{survfitJM()} or \code{predict()} and a vector of
posterior model weights, and returns a single object of the same class as the input 
objects with the combined predictions. We illustrate how this function can be used to 
produce the BMA prediction of $\pi_j(u \mid t)$ using the first five measurements of 
Patient~2 from the PBC dataset based on joint models \code{jointFit.pbc1}, 
\code{jointFit.pbc12}, \code{jointFit.pbc13}, \code{jointFit.pbc14}, and 
\code{jointFit.pbc15}. We start by computing the posterior model weights. As seen above
for the calculation these weights we need to compute the marginal densities 
$p(\mathcal D_n \mid M_k)$ and $p(\mathcal D_j(t) \mid M_k)$. The former is obtained using
the \code{logLik()} method for \code{JMbayes} objects, and the latter using function 
\code{marglogLik()}. The following code illustrates how this can be achieved:
\begin{Schunk}
\begin{Sinput}
R> Models <- list(jointFit.pbc1, jointFit.pbc12, jointFit.pbc13, 
+                 jointFit.pbc14, jointFit.pbc15)
\end{Sinput}
\end{Schunk}
\begin{Schunk}
\begin{Sinput}
R> log.p.Dj.Mk <- sapply(Models, marglogLik, newdata = ND[1:5, ])
R> log.p.Dn.Mk <- sapply(Models, logLik, marginal.thetas = TRUE)
R> log.p.Mk <- log(rep(1/length(Models), length(Models)))
\end{Sinput}
\end{Schunk}
Argument \code{newdata} of \code{marglogLik()} is used to provide the available data 
$\mathcal D_j(t)$ of the $j$-th subject, whereas argument \code{marginal.thetas} is invoked
in order the \code{logLik()} method to compute the marginal log-likelihood. As just 
mentioned, we should stress that \code{marglogLik()} and \code{logLik()} compute 
$\log p(\mathcal D_j(t) \mid M_k)$ and $\log p(\mathcal D_n \mid M_k)$, respectively. 
Hence, to calculate the weights we need to transform them back to the original scale, i.e.,
\begin{Schunk}
\begin{Sinput}
R> weightsBMA <- log.p.Dj.Mk + log.p.Dn.Mk + log.p.Mk
R> weightsBMA <- exp(weightsBMA - mean(weightsBMA))
R> weightsBMA <- weightsBMA / sum(weightsBMA)
\end{Sinput}
\end{Schunk}
Following we calculate the conditional survival probabilities based on each model, using
\code{survfitJM()}
\begin{Schunk}
\begin{Sinput}
R> survPreds <- lapply(Models, survfitJM, newdata = ND[1:5, ])
\end{Sinput}
\end{Schunk}
and finally we combine them using the call to \code{bma.combine()}: 
\begin{Schunk}
\begin{Sinput}
R> survPreds.BMA <- bma.combine(JMlis = survPreds, weights = weightsBMA)
R> survPreds.BMA
\end{Sinput}
\begin{Soutput}
Prediction of Conditional Probabilities for Event
	based on 200 Monte Carlo samples

$`2`
     times   Mean Median  Lower  Upper
1   4.9009 1.0000 1.0000 1.0000 1.0000
1   5.0072 0.9937 0.9942 0.9825 0.9987
2   5.3336 0.9728 0.9757 0.9226 0.9947
3   5.6599 0.9493 0.9558 0.8562 0.9910
4   5.9862 0.9231 0.9344 0.7831 0.9874
5   6.3126 0.8939 0.9116 0.7042 0.9841
6   6.6389 0.8617 0.8856 0.6201 0.9809
7   6.9652 0.8266 0.8587 0.5332 0.9780
8   7.2916 0.7892 0.8265 0.4464 0.9752
9   7.6179 0.7500 0.7950 0.3286 0.9725
10  7.9442 0.7096 0.7598 0.2026 0.9694
11  8.2706 0.6685 0.7187 0.1019 0.9652
12  8.5969 0.6275 0.6910 0.0384 0.9627
13  8.9232 0.5875 0.6550 0.0097 0.9614
14  9.2496 0.5494 0.6149 0.0016 0.9603
15  9.5759 0.5140 0.5823 0.0003 0.9583
16  9.9022 0.4817 0.5374 0.0001 0.9562
17 10.2286 0.4527 0.4809 0.0000 0.9543
18 10.5549 0.4268 0.4418 0.0000 0.9526
19 10.8812 0.4036 0.4067 0.0000 0.9510
20 11.2076 0.3828 0.3727 0.0000 0.9496
\end{Soutput}
\end{Schunk}


\subsection[Predictive accuracy]{Predictive accuracy}\label{Sec:DynPred-PredAcc}
The assessment of the predictive performance of time-to-event models has received a lot of
attention in the statistical literature. In general two main lines have emerged, namely 
one focusing on calibration, i.e., how well the model predicts the observed data 
\citep{schemper.henderson:00, gerds.schumacher:06} and a second on focusing on 
discrimination, i.e., how well can the model discriminate between patients that had the 
event from patients that did not \citep{harrell.et.al:96, pencina.et.al:08}. In the 
following we present discrimination and calibration measures suitably adapted to the 
dynamic prediction setting and their implementation in \pkg{JMbayes}.


\subsubsection[Discrimination]{Discrimination}
To measure the discriminative capability of a longitudinal marker we focus on a time 
interval of medical relevance within which the occurrence of events is of interest. 
In this setting, a useful property of the model would be to successfully discriminate 
between patients who are going to experience the event within this time frame from 
patients who will not. To put this formally, as before, we assume that we have collected 
longitudinal measurements $\mathcal Y_j(t) = \{ y_j(t_{jl});
0 \leq t_{jl} \leq t, l = 1, \ldots, n_j \}$ up to time point $t$ for subject $j$. We are 
interested in events occurring in the medically-relevant time frame $(t, t + \Delta t]$ 
within which the physician can take an action to improve the survival chance of the 
patient. Under the assumed model and the methodology presented in 
Section~\ref{Sec:DynPred-Est}, we can define a prediction rule using 
$\pi_j(t + \Delta t \mid t)$ that takes into account the available longitudinal 
measurements $\mathcal Y_j(t)$. In particular, for any value $c$ in $[0, 1]$ we can term 
subject $j$ as a case if $\pi_j(t + \Delta t \mid t) \leq c$ (i.e., occurrence of the 
event) and analogously as a control if $\pi_j(t + \Delta t \mid t) > c$. Thus, in this 
context, we define sensitivity and specificity as
\[
\Pr \bigl \{ \pi_j(t + \Delta t \mid t) \leq c \mid T_j^* \in (t, t + \Delta t] \bigr \},
\]
and
\[
\Pr \bigl \{ \pi_j(t + \Delta t \mid t) > c \mid T_j^* > t + \Delta t \bigr \},
\]
respectively. For a randomly chosen pair of subjects $\{i, j\}$, in which both subjects 
have provided measurements up to time $t$, the discriminative capability of the assumed 
model can be assessed by the area under the receiver operating characteristic curve (AUC),
which is obtained for varying $c$ and equals,
\[
\mbox{AUC}(t, \Delta t) = \Pr \bigl [ \pi_i(t + \Delta t \mid t) < \pi_j(t + \Delta t 
\mid t) \mid \{ T_i^* \in (t, t + \Delta t] \} \cap \{ T_j^* > t + \Delta t \} \bigr ], 
\]
that is, if subject $i$ experiences the event within the relevant time frame whereas 
subject $j$ does not, then we would expect the assumed model to assign higher probability 
of surviving longer than $t + \Delta t$ for the subject who did not experience the event. 
To summarize the discriminating power of the assumed model over the whole follow-up 
period, we need to take into account that the number of subjects contributing to the 
comparison of the fitted $\pi_i(t + \Delta t \mid t)$ with the observed data is not the 
same for all time points $t$. Following an approach similar to \citet{antolini.et.al:05} 
and \citet{heagerty.zheng:05}, we can utilize a weighted average of AUCs, i.e.,
\begin{equation}
\mbox{C}_{dyn}^{\Delta t} = \int_0^\infty \mbox{AUC}(t, \Delta t) \, \Pr \{ 
\mathcal E(t) \} \; dt \Big / \int_0^\infty \Pr \{ \mathcal E(t) \} \; dt, \label{Eq:dynC}
\end{equation}
where $\mathcal E(t) = \bigl [ \{ T_i^* \in (t, t + \Delta t] \} \cap \{ T_j^* > t + 
\Delta t \} \bigr ]$, and $\Pr \{ \mathcal E(t) \}$ denotes the probability that a random 
pair is comparable at $t$. We can call $\mbox{C}_{dyn}^{\Delta t}$ a dynamic concordance 
index since it summarizes the concordance probabilities over the follow-up period. Note 
also that $\mbox{AUC}(t, \Delta t)$ and as a result also $\mbox{C}_{dyn}^{\Delta t}$ 
depend on the length $\Delta t$ of the time interval of interest, which implies that 
different models may exhibit different discrimination power for different $\Delta t$.

For the estimation of $\mbox{AUC}(t, \Delta t)$ and $\mbox{C}_{dyn}^{\Delta t}$ we need to 
take care of two issues, namely, the calculation of the integrals in the definition of 
\eqref{Eq:dynC} and censoring. For the former we use the 15-point Gauss-Kronrod quadrature 
rule. Estimation of $\mbox{AUC}(t, \Delta t)$ is directly based on its definition, namely
by appropriately counting the concordant pairs of subjects. More specifically, we have
\[
\mbox{\aucHat}(t, \Delta t) = \mbox{\aucHat}_1(t, \Delta t) + 
\mbox{\aucHat}_2(t, \Delta t).
\]
$\mbox{\aucHat}_1(t, \Delta t)$ refers to the pairs of subjects who are comparable (i.e,
their observed event times can be ordered),
\begin{eqnarray*}
\Omega_{ij}^{(1)}(t) & = & \bigl [\{T_i \in (t, t + \Delta t] \} \cap \{\delta_i = 1 \} 
\bigr ] \cap \{ T_j > t + \Delta t \},
\end{eqnarray*}
where $i, j = 1, \ldots, n$ with $i \neq j$. For such comparable subjects $i$ and $j$, we 
can estimate and compare their survival probabilities $\pi_i(t + \Delta t \mid t)$ and 
$\pi_j(t + \Delta t \mid t)$, based on the methodology presented in 
Section~\ref{Sec:DynPred-Est}. This leads to a natural estimator for 
$\mbox{AUC}_1(t, \Delta t)$ as the proportion of concordant subjects out of the 
set of comparable subjects at time $t$:
\[
\mbox{\aucHat}_1(t, \Delta t) = \frac{\sum_{i = 1}^n \sum_{j = 1; j \neq i}^n I \{ \hat 
\pi_i(t + \Delta t \mid t) < \hat \pi_j(t + \Delta t \mid t) \} \times I 
\{\Omega_{ij}^{(1)}(t)\}}{\sum_{i=1}^n \sum_{j=1; j \neq i}^n I\{\Omega_{ij}^{(1)}(t)\}},
\]
where $I(\cdot)$ denotes the indicator function. Analogously, 
$\mbox{\aucHat}_2(t, \Delta t)$ refers to the pairs of subjects who due to censoring 
cannot be compared, namely
\[
\Omega_{ij}^{(2)}(t) = \bigl [\{T_i \in (t, t + \Delta t] \} \cap \{\delta_i = 0 \} 
\bigr ] \cap \{ T_j > t + \Delta t \},
\]
with again $i, j = 1, \ldots, n$ with $i \neq j$. Concordant subjects in this set 
contribute to the overall AUC appropriately weighted with the probability that they would 
be comparable, i.e.,
\[
\mbox{\aucHat}_2(t, \Delta t) = \frac{\sum_{i = 1}^n \sum_{j = 1; j \neq i}^n I \{ \hat 
\pi_i(t + \Delta t \mid t) < \hat \pi_j(t + \Delta t \mid t) \} \times I 
\{\Omega_{ij}^{(2)}(t)\} \times \hat \nu_i(t + \Delta t \mid T_i)}{\sum_{i=1}^n 
\sum_{j=1; j \neq i}^n I\{\Omega_{ij}^{(2)}(t)\} \times \hat \nu_i(t + \Delta t \mid T_i)},
\]
with $\hat \nu_i(t + \Delta t \mid T_i) = 1 - \hat \pi_i(t + \Delta t \mid T_i)$ being the 
probability that subject $i$ who survived up to time $T_i$ will have the event before 
$t + \Delta t$. 

Having estimated $\mbox{AUC}(t, \Delta t)$, the next step in estimating 
$\mbox{C}_{dyn}^{\Delta t}$ is to obtain estimates for the weights 
$\mbox{Pr} \{ \mathcal E(t) \}$. We observe that these can be rewritten as
\begin{eqnarray*}
\Pr \{ \mathcal E(t) \} & = & \Pr \bigl [ \{ T_i^* \in (t, t + \Delta t] \} \cap 
\{ T_j^* > t + \Delta t \} \bigr ] \\
& = & \Pr (T_i^* \in (t, t + \Delta t]) \times \Pr( T_j^* > t + \Delta t)\\
& = & \bigl \{ S(t) - S(t + \Delta t) \bigr \} S(t + \Delta t),
\end{eqnarray*}
where the simplification in the second line comes from the independence of subjects $i$ 
and $j$, and $S(\cdot)$ here denotes the marginal survival function. In practice 
calculation of $\mbox{C}_{dyn}^{\Delta t}$ is restricted into a follow-up 
interval $[0, t_{max}]$ where we have information. Let $t_1, \ldots, t_{15}$ denote the 
re-scaled abscissas of the Gauss-Kronrod rule in the interval $[0, t_{max}]$ with 
corresponding weights $\varpi_1, \ldots, \varpi_{15}$. We combine the estimates 
$\mbox{\aucHat}(t_k, \Delta t)$, $k = 1, \ldots, 15$ with the estimates of the weights 
$\Pr \{ \mathcal E(t) \}$ to obtain
\[
\widehat{\mbox{C}}_{dyn}^{\Delta t} = \frac{\sum_{k = 1}^{15} \varpi_k \mbox{\aucHat}(t_k,
\Delta t) \times \widehat{\mbox{Pr}} \{ \mathcal E(t_k) \}}{\sum_{k = 1}^{15} \varpi_k 
\widehat{\mbox{Pr}} \{ \mathcal E(t_k) \}},
\]
where $\widehat{\Pr} \{ \mathcal E(t_k) \} = \bigl \{ \widehat S(t_k) - \widehat 
S(t_k + \Delta t) \bigr \} \widehat S (t_k + \Delta t)$, with $\widehat S (\cdot)$ 
denoting here the Kaplan-Meier estimate of the marginal survival function $S(\cdot)$.

The $\mbox{AUC}(t, \Delta t)$ and the dynamic discrimination index can be calculated for
joint models fitted by \code{jointModelBayes()} using functions \code{aucJM()} and 
\code{dynCJM()}, respectively. We illustrate their use based again on joint model 
\code{jointFit.pbc15}. The basic call to \code{aucJM()} requires the user to provide the
fitted joint model object, the data frame upon which the AUC is to be calculated, the time
point $t$ (argument \code{Tstart}) up to which longitudinal measurements are to be used 
and the length of the time window $\Delta t$ (argument \code{Dt})\footnote{instead of 
giving \code{Dt} the user may choose to directly give the horizon time $t + \Delta t$ in 
the argument \code{Thoriz}.}:
\begin{Schunk}
\begin{Sinput}
R> auc.pbc15 <- aucJM(jointFit.pbc15, newdata = pbc2, Tstart = 5, Dt = 2)
R> auc.pbc15
\end{Sinput}
\end{Schunk}
\begin{Schunk}
\begin{Soutput}
	Time-dependent AUC for the Joint Model jointFit.pbc15

Estimated AUC: 0.842
At time: 7
Using information up to time: 5 (202 subjects still at risk)
\end{Soutput}
\end{Schunk}
We observe that using the first five year longitudinal measurements, serum bilirubin 
exhibits nice discrimination capabilities for patients who are to die within a two-year 
time frame. To investigate if this is also the case during the whole follow-up period, we
calculate the dynamic discrimination index for the same time window. The syntax of 
\code{dynCJM()} is (almost) identical to the one of \code{aucJM()}, i.e.,
\begin{Schunk}
\begin{Sinput}
R> dynC.pbc15 <- dynCJM(jointFit.pbc15, newdata = pbc2, Dt = 2)
R> dynC.pbc15
\end{Sinput}
\end{Schunk}
\begin{Schunk}
\begin{Soutput}
	Dynamic Discrimination Index for the Joint Model jointFit.pbc15

Estimated dynC: 0.8496
In the time interval: [0, 14.3057]
Length of time interval: 2
\end{Soutput}
\end{Schunk}
The estimate of $\mbox{C}_{dyn}^{\Delta t=2}$ is almost identical to the one of
$\mbox{AUC}(t = 5, \Delta t = 2)$ indicating that serum bilirubin can discriminate well 
between patients during follow-up.


\subsubsection[Prediction error]{Prediction error}
The assessment of the accuracy of predictions of survival models is typically based on the
expected error of predicting future events. In our setting, and again taking into account 
the dynamic nature of the longitudinal outcome, it is of interest to predict the 
occurrence of events at $u > t$ given the information we have recorded up to time $t$. 
This gives rise to expected prediction error:
\[
\mbox{PE}(u \mid t) = E \bigl [ L\{N_i(u) - \pi_i(u \mid t)\} \bigr ],
\]
where $N_i(t) = I(T_i^* > t)$ is the event status at time $t$, $L(\cdot)$ denotes a loss 
function, such as the absolute or square loss, and the expectation is taken with respect 
to the distribution of the event times. An estimate of $\mbox{PE}(u \mid t)$ that accounts
for censoring has been proposed by \citet{henderson.et.al:02}:
\begin{eqnarray*}
\lefteqn{\nonumber \widehat{\mbox{PE}}(u \mid t) = \{n(t)\}^{-1} \sum_{i: T_i \geq t} 
I(T_i \geq u) L\{1 - \hat \pi_i(u \mid t)\} + \delta_i I(T_i < u) L\{0 - 
\hat \pi_i(u \mid t)\}}\\
&& + (1 - \delta_i) I(T_i < u) \Bigl [ \hat \pi_i(u \mid T_i) L\{1 - 
\hat \pi_i(u \mid t)\} + \{1 - \hat \pi_i(u \mid T_i)\} L\{0 - 
\hat \pi_i(u \mid t)\} \Bigr ],
\end{eqnarray*}
where $n(t)$ denotes the number of subjects at risk at time $t$. The first two terms in 
the sum correspond to patients who were alive after time $u$ and dead before $u$, 
respectively; the third term corresponds to patients who were censored in the interval 
$[t, u]$. Using the longitudinal information up to time $t$, $\mbox{PE}(u \mid t)$ 
measures the predictive accuracy at the specific time point $u$. Alternatively, we could 
summarize the error of prediction in a specific interval of interest, say $[t, u]$, by 
calculating a weighted average of $\{\mbox{PE}(s \mid t), t < s< u\}$ that corrects for 
censoring, similarly to $\mbox{C}_{dyn}^{\Delta t}$. An estimator of this type for the 
integrated prediction error has been suggested by \citet{schemper.henderson:00}, which 
adapted to our time-dynamic setting takes the form
\[
\mbox{\ipeHat}(u \mid t) = \frac{\sum_{i: t \leq T_i \leq u} \delta_i \bigl 
\{ \widehat S_C(t) / \widehat S_C(T_i) \bigr \} \widehat{\mbox{PE}}(T_i \mid t)}{
\sum_{i: t \leq T_i \leq u} \delta_i \bigl \{ \widehat S_C(t) / \widehat S_C(T_i) 
\bigr \}},
\]
where $\widehat S_C(\cdot)$ denotes the Kaplan-Meier estimator of the censoring time 
distribution.

Both PE and IPE can be calculated for joint models fitted by \code{jointModelBayes()} 
using function \code{prederrJM()}. This has a similar syntax as function \code{aucJM()},
and requires a fitted joint model, a data frame based on which the prediction error will 
be calculated, and the time points $t$ (argument \code{Tstart}) and $u$ 
(argument \code{Thoriz}) that denotes up to which time point to use the longitudinal 
information and at which time point to make the prediction, respectively. For model 
\code{jointFit.pbc15} using the biomarker information during  the first five years of 
follow-up the estimated prediction error at year seven is
\begin{Schunk}
\begin{Sinput}
R> pe.pbc15 <- prederrJM(jointFit.pbc15, pbc2, Tstart = 5, Thoriz = 7)
R> pe.pbc15
\end{Sinput}
\end{Schunk}
\begin{Schunk}
\begin{Soutput}
Prediction Error for the Joint Model jointFit.pbc15

Estimated prediction error: 0.107
At time: 7
Using information up to time: 5 (202 subjects still at risk)
Loss function: square
\end{Soutput}
\end{Schunk}
By default the loss function is the square one (i.e., $L(x) = x^2$), but the user may 
specify the absolute loss or define her own loss function using argument \code{lossFun}.
The integrated prediction error can be simply calculated by setting logical argument 
\code{interval} to \code{TRUE} in the call to \code{ prederrJM()}; for example, for the 
same joint model and in the interval $[5, 9]$ the IPE is calculated with the code:
\begin{Schunk}
\begin{Sinput}
R> ipe.pbc15 <- prederrJM(jointFit.pbc15, pbc2, Tstart = 5, 
+                         Thoriz = 9, interval = TRUE)
R> ipe.pbc15
\end{Sinput}
\end{Schunk}
\begin{Schunk}
\begin{Soutput}
Prediction Error for the Joint Model jointFit.pbc15

Estimated prediction error: 0.0907
In the time interval: [5, 9]
Using information up to time: 5 (202 subjects still at risk)
Loss function: square
\end{Soutput}
\end{Schunk}


\subsubsection[Validation]{Validation}
In the previous sections we have seen how the predictive performance of model 
\code{jointFit.pbc15} can be assessed in terms of discrimination and calibration on the 
PBC dataset. However, as it is know from the prognostic models literature 
\citep[see e.g.,][]{harrell:01}, these estimates of predictive performance may be 
over-optimistic because they do not account for the fact the model was also fitted in the 
same dataset. One standard approach to obtain better, more objective, estimates of 
predictive ability is to utilize the cross-validation technique. The following code 
illustrates how we could implement 10-fold cross-validation using package \pkg{parallel}. 
First, we load the package and create 10 random splittings of the PBC dataset:
\begin{Schunk}
\begin{Sinput}
R> library("parallel")
R> set.seed(123)
R> V <- 10
R> n <- nrow(pbc2.id)
R> splits <- split(seq_len(n), sample(rep(seq_len(V), length.out = n)))
\end{Sinput}
\end{Schunk}
Following we define a function that takes as argument the above defined splittings, 
creates the training and testing datasets, fits joint model \code{jointFit.pbc15} in the
training dataset, and calculates the AUC and the PE in the test dataset:
\begin{Schunk}
\begin{Sinput}
R> CrossValJM <- function (i) {
+      library("JMbayes")
+      pbc2$status2 <- as.numeric(pbc2$status != "alive")
+      pbc2.id$status2 <- as.numeric(pbc2.id$status != "alive")
+      
+      trainingData <- pbc2[!pbc2$id %in% i, ]
+      trainingData.id <- trainingData[!duplicated(trainingData$id), ]
+      testingData <- pbc2[pbc2$id %in% i, ]
+      
+      lmeFit.pbc1 <- lme(log(serBilir) ~ ns(year, 2), data = trainingData,
+                         random = ~ ns(year, 2) | id)
+      coxFit.pbc1 <- coxph(Surv(years, status2) ~ drug * age, 
+                           data = trainingData.id, x = TRUE)
+      
+      dForm <- list(fixed = ~ 0 + dns(year, 2), random = ~ 0 + dns(year, 2), 
+                    indFixed = 2:3, indRandom = 2:3)
+      tf1 <- function (x, data) {
+          cbind(x, "^2" = x*x)
+      }
+      tf2 <- function (x, data) {
+          cbind(x, "drugD-penicil" = x * (data$drug == 'D-penicil'))
+      }
+      jointFit.pbc15 <- 
+          jointModelBayes(lmeFit.pbc1, coxFit.pbc1, timeVar = "year",
+                          param = "td-both", extraForm = dForm, 
+                          transFun = list(value = tf1, extra = tf2))
+      
+      auc <- aucJM(jointFit.pbc15, newdata = testingData, 
+                   Tstart = 5, Thoriz = 7)
+      pe <- prederrJM(jointFit.pbc15, newdata = testingData, 
+                      Tstart = 5, Thoriz = 7)    
+      list(auc = auc, pe = pe)
+  }
\end{Sinput}
\end{Schunk}
We run function \code{CrossValJM()} in parallel using five processors/cores by first 
creating the corresponding cluster and then using \code{parLapply()}:
\begin{Schunk}
\begin{Sinput}
R> cl <- makeCluster(5)
R> res <- parLapply(cl, splits, CrossValJM)
R> stopCluster(cl)
\end{Sinput}
\end{Schunk}
The averaged AUCs and PEs from the 10 random splits of the PBC dataset are calculated with
the code:
\begin{Schunk}
\begin{Sinput}
R> mean(sapply(res, function (x) x$auc$auc))
\end{Sinput}
\begin{Soutput}
[1] 0.8400467
\end{Soutput}
\begin{Sinput}
R> mean(sapply(res, function (x) x$pe$prederr))
\end{Sinput}
\begin{Soutput}
[1] 0.1230112
\end{Soutput}
\end{Schunk}
We observe that the cross-validated estimate of the AUC is identical the one obtained in 
the original dataset, whereas for the prediction error there is a slight over-optimism.

\newpage
\section[Future plans]{Future plans} \label{Sec:FuturePlans}
In this paper we have illustrated the capabilities of package \pkg{JMbayes} for fitting 
joint models for longitudinal and time-to-event data under a Bayesian approach. As we have
seen, the current version of the package provides several options for fitting different 
types of joint models, but nonetheless several extensions are planned in the future to 
further expand on what is currently available. These include among others:
\begin{itemize}
\item The consideration of multiple longitudinal outcomes, while allowing for the various
association structures we have presented in Sections~\ref{Sec:JMbayesBasicUse-Ass} and 
\ref{Sec:JMbayesBasicUse-Transf}.

\item Handling of exogenous time-varying covariates by supplying a time-dependent Cox model
as an argument to \code{jointModelBayes()}.

\item Extend functionality in the survival submodel to handle, competing risks, recurrent 
events, and left- and interval-censored event time data.

\item Update dynamic predictions to handle the aforementioned extensions.
\end{itemize}


\bibliography{JMbayes}


\newpage
\appendix
\section[MCMC diagnostic plots]{MCMC diagnostic plots} \label{Sec:App-DiagnMCMC}
The \code{plot()} method for objects produced by \code{jointModelBayes()} produces 
diagnostic plots for the MCMC, namely trace, auto-correlation and kernel density estimated 
plots. In addition, the \code{plot()} method can be used to create the figure of the CPO. 
As an example, we produce trace and density plots for the joint model \code{jointFit.pbc1}
that was fitted in Section~\ref{Sec:JMbayesBasicUse-Std}. To avoid lengthy output we just
illustrate how these plots are produced for the parameters of the longitudinal submodel.
The relevant code is:
\begin{Schunk}
\begin{Sinput}
R> plot(jointFit.pbc1, param = c("betas", "sigma", "D"))
R> plot(jointFit.pbc1, which = "density", param = c("betas", "sigma", "D"))
\end{Sinput}
\end{Schunk}
\begin{figure}
\centering
\includegraphics[width = 0.9\textwidth]{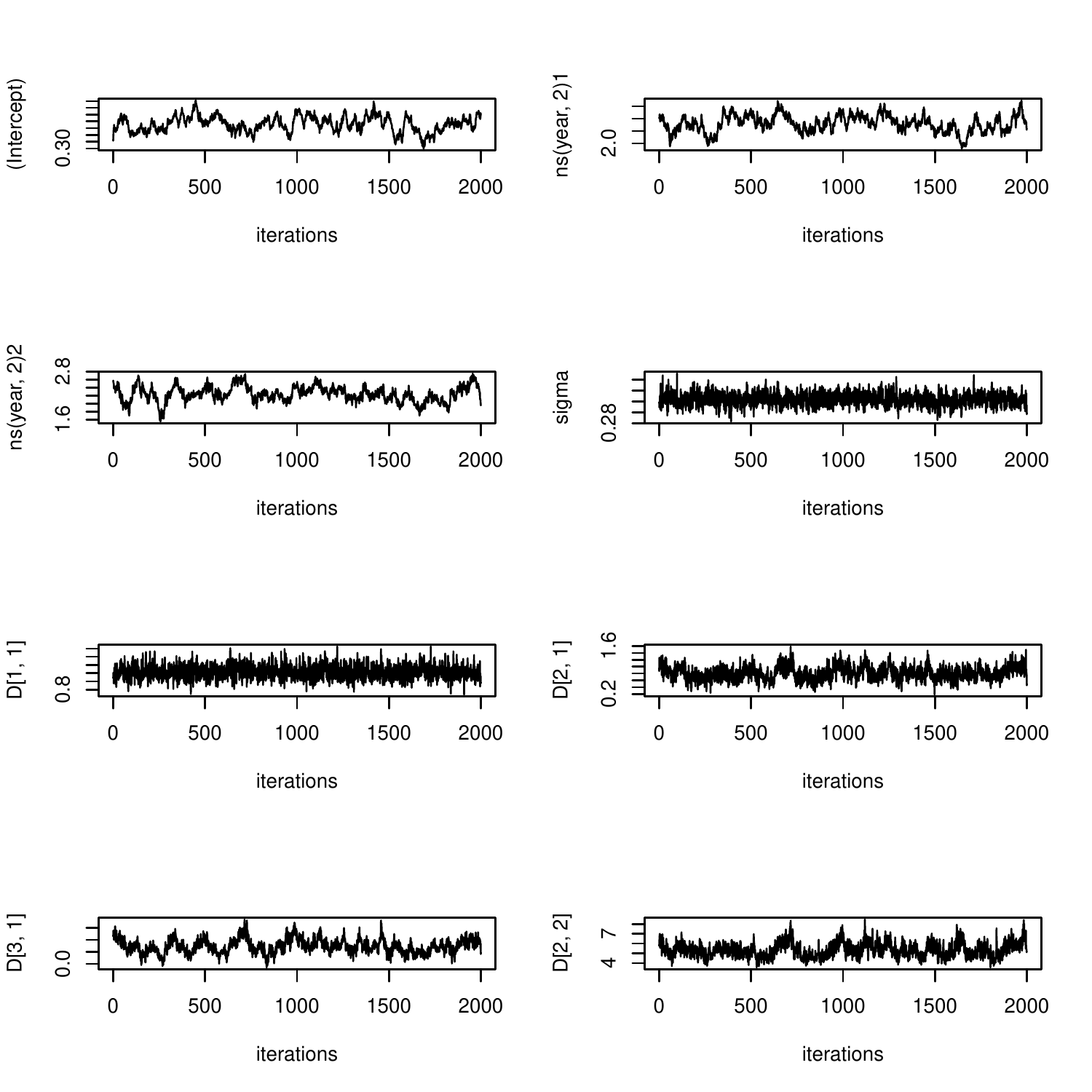}
\caption{Traceplots for the parameters of the longitudinal submodel from 
\code{jointFit.pbc1}.} \label{Fig:MCMCdiagnTrace}
\end{figure}
\begin{figure}
\centering
\includegraphics[width = 0.9\textwidth]{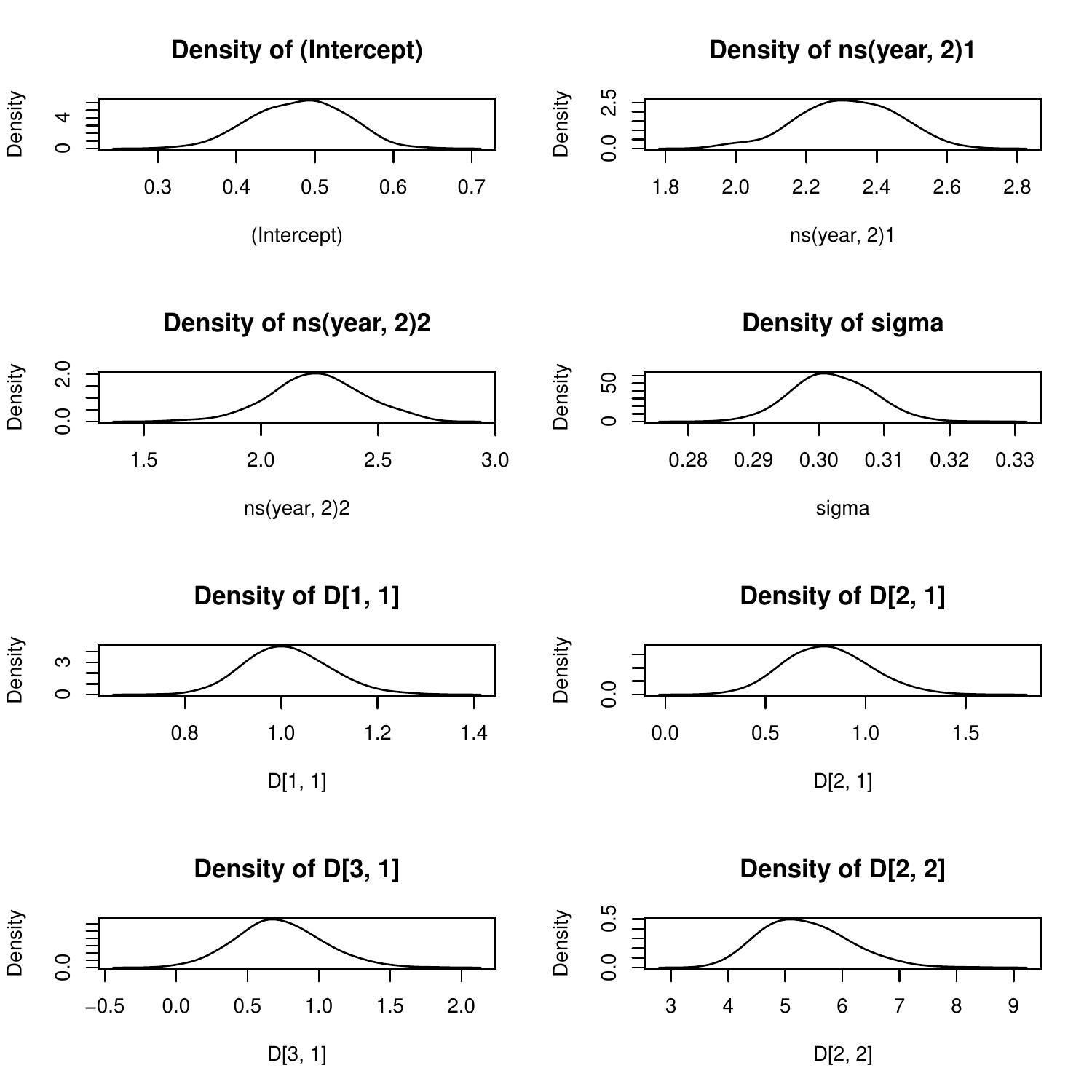}
\caption{Kernel density estimation plots for the parameters of the longitudinal submodel 
from \code{jointFit.pbc1}.} \label{Fig:MCMCdiagnDens}
\end{figure}

\end{document}